\documentclass[twocolumn,aps,superscriptaddress,nofootinbib,floatfix,linenumbers]{revtex4}
\usepackage{bm}
\usepackage{bbm}
\usepackage{graphicx} 
\usepackage[utf8]{inputenc}
\usepackage[dvipsnames]{xcolor}
\usepackage{amsmath}
\usepackage{amssymb}
\usepackage{lipsum}
\usepackage{ulem}
\usepackage{float}
\usepackage{amsfonts}
\usepackage{physics}
\usepackage{slashed}
\usepackage{wasysym}
\usepackage[hidelinks]{hyperref}
\usepackage{makecell}
\usepackage[thinlines]{easytable}
\usepackage[caption=false]{subfig}
\usepackage{tabularray}
\NewColumnType{M}[1]{Q[m,c,#1]} 
\usepackage{breakcites}

\usepackage{feynmp-auto} 
\usepackage{comment}

\begin{document}

\title{Shear viscosity of a binary mixture for a relativistic fluid at high temperature}

\author{Gabriele Parisi}
\email{gabriele.parisi@dfa.unict.it}
\affiliation{Department of Physics and Astronomy, University of Catania, Via S. Sofia 64, I-95125 Catania}
\affiliation{INFN-Laboratori Nazionali del Sud, Via S. Sofia 62, I-95123 Catania, Italy}

\author{Vincenzo Nugara}
\email{vincenzo.nugara@phd.unict.it}
\affiliation{Department of Physics and Astronomy, University of Catania, Via S. Sofia 64, I-95125 Catania}
\affiliation{INFN-Laboratori Nazionali del Sud, Via S. Sofia 62, I-95123 Catania, Italy}

\author{Salvatore Plumari}
\email{salvatore.plumari@dfa.unict.it}
\affiliation{Department of Physics and Astronomy, University of Catania, Via S. Sofia 64, I-95125 Catania}
\affiliation{INFN-Laboratori Nazionali del Sud, Via S. Sofia 62, I-95123 Catania, Italy}

\author{Vincenzo Greco}
\email{greco@lns.infn.it}
\affiliation{Department of Physics and Astronomy, University of Catania, Via S. Sofia 64, I-95125 Catania}
\affiliation{INFN-Laboratori Nazionali del Sud, Via S. Sofia 62, I-95123 Catania, Italy}

\begin{abstract}
The determination of the shear viscosity is a central topic in various areas of modern physics. In particular, it is often necessary to evaluate the shear viscosity $\eta$ of fluids made up of more than one species, all interacting with different cross sections. Since it may be difficult to extract information on the interaction among different species, various combinations of the viscosities of the individual components are often used. We work in the Chapman-Enskog framework and investigate on binary mixtures, by comparing such single component combinations with a full 2-component formalism: we find that, in most cases, the full viscosity is well approximated by a weighted linear average of the single component viscosities, although this result is far from being general. Moreover, we validate our 2-component Chapman-Enskog results for $\eta$ by comparing them with an independent numerical simulation of the Boltzmann equation, which estimates the shear viscosity via a Green-Kubo formula, in the case of a quasi-particle system that reproduces lattice QCD thermodynamics. We see that the temperature dependence of $\eta/s$ of such system of quarks and gluons is not well described by combinations of the individual components, highlighting the importance of inter-species scattering.
\end{abstract}

\pacs{}
\keywords{}
\maketitle

\section{Introduction}

Shear viscosity is a fundamental transport coefficient that quantifies the resistance of a fluid to deformations under shear stress. It is a key property of a fluid, since it characterizes the internal friction arising when adjacent layers of a fluid move relative to one another, so that a force applied on the upper layer is transferred to the next, with a small defect due to friction \cite{Landau}. In this sense, it represents a measure of its dissipative behavior. 

Shear viscosity plays a central role in the hydrodynamic description of many physical systems, governing the rate at which momentum is transported across layers of the fluid. As such, it controls the approach towards local equilibrium and can influence macroscopic observables for different physical systems. In classical fluids, such as water or air, viscosity determines energy dissipation, flow behavior and the onset of turbulence. In astrophysics, shear viscosity plays a crucial role in the damping of stellar oscillations and in the cooling of neutron stars, arising mainly from scattering processes among the constituents of dense nuclear matter \cite{Shternin:2008es,Zhang:2010jf,Jaccarino:2012zz}.
In condensed matter physics it governs transport in liquid metals, ultracold atomic gases, and superfluid phases \cite{ bruun2005viscosity,Rupak:2007vp,enss2011viscosity,levchenko2020transport}. Shear viscosity is an important concept also in studies of superfluidity and of the low-temperature behavior of cold Fermi gases at unitarity \cite{Manuel:2011ed,Manuel:2012rd,Mannarelli:2012eg,Mannarelli:2012su}.

In high-energy physics, the concept extends to exotic states of matter like the Quark-Gluon Plasma (QGP) \cite{Noronha-Hostler:2012ycm,El:2012cr, Zaccone:2023ozw, Zaccone:2024zel}, where experimental observations suggest that the QGP exhibits nearly perfect fluidity with an exceptionally low shear viscosity to entropy density ratio ($\eta/s$), approaching the conjectured Kovtun–Son–Starinets (KSS) bound of $\eta/s=1/4\pi$ \cite{Kovtun:2004de}. Understanding shear viscosity is therefore essential for modeling both conventional fluid dynamics and strongly interacting quantum systems. It serves as a key parameter in the study of turbulence, thermalization, and transport phenomena across a broad range of physical contexts.\\

Despite the key role of shear viscosity, its evaluation becomes particularly challenging when the medium is composed of more than one species.  In a multi-component fluid one has to take into account for both intra-species and inter-species interactions, each of which may contribute differently to transport properties. In particular, a relevant question arises when one has to evaluate the shear viscosity of a mixture of two different species, without detailed information on the interaction
among one species and the other.  For this purpose, in the literature many approaches consider the shear viscosity $\eta$ of a binary mixture as a combination of the two shear viscosities $\eta_1$ and $\eta_2$ of the single components. This method has been widely employed in the study of cosmic objects (stellar plasmas \cite{Haxhimali:2015}, superfluidity \cite{Trabucco:2025duz}, neutron stars \cite{Shternin:2008es,Zhang:2010jf,Jaccarino:2012zz}) and also in condensed matter physics \cite{ARNAULT2013711}. Just as an example, one can evaluate the shear viscosity of a binary mixture by noting that the mean free path of the two 2-component fluid is $\lambda= \lambda_1+  \lambda_2$, where $\lambda_{1,2}$ are the mean free paths of the two components. Since for the viscosity $\eta\propto\lambda^{-1}$, this may suggest that $1/\eta=1/\eta_1+1/\eta_2$ \cite{Manuel:2012rd,Mannarelli:2012eg,Mannarelli:2012su}. 

In any case, whatever combination of $\eta_1$ and $\eta_2$ we intend to use to describe the viscosity of the mixture, the result will contain no information about the interaction among the two species, encoded by the differential cross section $\sigma_{12}$. Depending on the magnitude of $\sigma_{12}$, its effect on the actual viscosity of the mixture may be huge, thus the regime of validity of any form of single-component interpolation may be questionable. 

The purpose of this work is to address this issue by studying the shear viscosity of binary mixtures within the Chapman--Enskog (CE) formalism, which provides a systematic framework to extract transport coefficients from the Boltzmann equation, based on an expansion of the phase space distribution function of the fluid around equilibrium.
We compare the two-component CE result for $\eta$ with various combinations of the shear viscosities of the single components $\eta_1$ and $\eta_2$ (evaluated using the single component CE formalism), in order to determine the domain of validity of such approximations. Finally, we will compare our findings for a realistic case with the results from a numerical simulation of the relativistic Boltzmann equation.\\

The paper is structured as follows. In Section \ref{sec:CE} we will introduce the Chapman-Enskog formulas for $\eta$, for both 1- and 2-component fluids. In Section \ref{Results_averages} we will compare various 1-component interpolations against the 2-component full result, highlighting the regime of validity of each of these approximations. In Section \ref{sec:GK} we will introduce the Green-Kubo method, which will provide an independent measure of $\eta$ based on the numerical solution of the Boltzmann-Vlasov equations. Finally, in Section \ref{Results for etaovers} we compare the Green-Kubo and the Chapman-Enskog results for the ratio $\eta/s$. We then report our Conclusions and Outlook.
This work also contains 5 Appendices: those contain technical details on the calculations of the paper, which have not been included in the main text for the sake of clarity.

Throughout this work we will use natural units of measure, i.e. we assume $\hbar=c=k_B=1$.


\section{Shear viscosity in Chapman-Enskog}
\label{sec:CE}
The Chapman-Enskog (CE) expansion is the conventional formalism used to derive fluid dynamic quantities from the Boltzmann equation.
It aims to obtain a solution of the transport equation valid in the hydrodynamic stage of equilibration, i.e. in the phase in which the spatial non-uniformities slowly disappear and the system relaxes to complete equilibrium \cite{DeGroot:1980dk}. It was originally developed for non-relativistic systems, but then Israel proved that it could be used with almost no modifications to describe relativistic systems as well \cite{Chapman_book_1974, Israel_1963}.

The CE formalism corresponds to the microscopic implementation of a gradient expansion, assuming that the single-particle distribution function depends only on the five primary fluid dynamical variables: temperature, chemical potential and the three independent components of the fluid velocity field (as well as their gradients). The series generated by the Chapman-Enskog procedure is generally thought to be asymptotic rather than convergent: this indicates that the first Chapman-Enskog approximation, which corresponds to linear laws for the transport phenomena, is the most significant.

In more detail, for a system not too far from equilibrium, one can express the Lorentz-invariant distribution function as follows \cite{DeGroot:1980dk, Wiranata:2012br, Prakash:1993bt}:
\begin{equation}
    f=f_0(1+\phi),
    \label{eq:distribution_function_CE}
\end{equation}
in which the function $\phi$ describes the 
deviation from the local equilibrium Boltzmann distribution function (i.e. $f_0$) and it is assumed to satisfy $|\phi|\ll 1$. The explicit form of $f_0$ is \cite{Prakash:1993bt}
\begin{equation}
    f_0=\frac{\rho z \exp\left(u_\alpha p^\alpha/T\right)}{4\pi m^3 K_2(z)}.
    \label{eq:boltzmann_distribution_eq}
\end{equation}
In Eq. \eqref{eq:boltzmann_distribution_eq}, $\rho\equiv \rho(x)$ and $T\equiv T(x)$ are the particle-number density and temperature in a proper coordinate system, $u\equiv u(x)$ is the four-velocity of the hydrodynamic particle flux (such that $u^\alpha u_\alpha=-1$) and $K_2(z)$ is the modified Bessel function of the second kind with $z\equiv m/T$. The corrections to the local distribution function (included in the term $\phi$) are systematically expressed as an expansion in powers of the Knudsen number \cite{Denicol:2012es}.

This formalism enables the calculation of the shear viscosity $\eta$. Furthermore, it provides flexibility in choosing both the order of approximation and the type of medium to be studied, whether it be a homogeneous gas or a multi-component mixture.

\subsection{Chapman-Enskog for homogeneous gases}
\label{CE homogeneous gas}
Following the approach outlined in \cite{Wiranata:2012br}, we can compute the shear viscosity $\eta$ within the CE approximation for the general case of relativistic particles with finite mass, colliding with a non-isotropic, energy-dependent cross section. \\
First, we focus on a single-component gas.
The first order approximation for the shear viscosity is given by:\footnote{For homogeneous gases also second and higher order approximations are available, cfr. \cite{Wiranata:2012br}. However, we checked that the second order approximation is very close to the first order result, therefore in the paper we will just stick to first order CE.}
\begin{align}
    \eta=\frac{1}{10}T\frac{\gamma_0^2}{c_{00}},\label{4.9}
\end{align}
where:
\begin{align}
    \gamma_0=&-10\hat{h},\nonumber\\
    c_{00}=&16\left(\omega_2^{(2)}-\frac{1}{z}\omega_1^{(2)}+\frac{1}{3z^2}\omega_0^{(2)}\right).
    \label{eq:CE_1comp_parameters}
\end{align}
In the above expressions $\hat{h}\equiv K_3(z)/K_2(z)$, being $K_n(z)$ the modified Bessel function of the second kind, whereas $\omega_i^{(s)}$ is the so-called relativistic omega integral \cite{Wiranata:2012br}:
\begin{align}
    \omega_i^{(s)}&=\frac{2\pi z^3}{K_2(z)^2}\int_1^\infty \dd y\, y^i(y^2-1)^3K_j(2zy)\cdot \label{4.11}\\
    &\int_0^\pi \dd\theta\sin\theta\,\frac{d\sigma}{d\Omega}(s,\theta)(1-\cos^s\theta).\nonumber
\end{align}
Here $j=5/2+(-1)^i/2$, $y=\sqrt{s}/2m$ and $d\sigma/d\Omega$ is the differential cross section, which is obtained from the (properly summed and averaged) squared scattering matrix as:
\begin{equation}
    \frac{d \sigma}{d\Omega}=\frac{1}{64\pi^2 s}|\mathcal{M}(\theta,\varphi)|^2,
    \label{19}
\end{equation}
being $s$ is the proper Mandelstam variable. Here note that the Chapman-Enskog approximation relies on elastic scattering only, hence in this model the particles retain their species after the collision.

\subsection{Chapman-Enskog for binary mixtures}
\label{binary mixture}
The Chapman-Enskog formalism allows for the calculation of the shear viscosity in binary mixtures, where the two species interact with distinct cross sections among themselves and with each other. This 2-component extension follows from a straightforward extension of Eq. \eqref{eq:distribution_function_CE}, see \cite{DeGroot:1980dk} for details. In this section we follow Ref. \cite{Wiranata:2013oaa}, in which the first-order CE approximation for $\eta$ in a binary mixture has been derived. While the formalism can, in principle, be generalized to mixtures with three or more components, such extensions introduce substantial algebraic complexity. For this reason, we restrict our analysis to the binary case, which already catches essential features of multi-component systems within the CE expansion.\\

At first order in the CE approximation, the shear viscosity of a binary mixture is given by:
\begin{equation}
    \eta=\frac{T}{10}\frac{\gamma_1^2c_{22}+\gamma_2^2c_{11}-2\gamma_1\gamma_2c_{12}}{c_{11}c_{22}-c_{12}^2},
    \label{two_comp_eta}
\end{equation}
where $\gamma_k=-10c_k[K_3(z_k)/K_2(z_k)]$, being $K_n(z_k)$ the modified Bessel function of order $n$ and $z_k=m_k/T$ ($k=1,2$ labels the particle species).\\
The coefficients $c_{kl}$ in \eqref{two_comp_eta} are given by:
\begin{align}
    c_{kk}&=c_k^2c_{00}+\tilde{c}_{kk},
    \label{1}\\
    c_{kl}&=\tilde{c}_{kl}~~~~\text{for }k\neq l.
    \label{2}
\end{align}
and the concentrations $c_k$ are defined as $c_k=\rho_k/\rho$, being $\rho_k$ the mass density of each species (mass times the number density) and $\rho=\sum_k \rho_k$ the total mass density.\\ The term $c_{00}=c_{00}(z_k)$ in \eqref{1} accounts for contributions from interactions between two identical particles of type $k$:
\begin{equation}
    c_{00}(z_k)=16[\omega_2^{(2)}(z_k)-\omega_1^{(2)}(z_k)/z_k+\omega_0^{(2)}(z_k)/(3z_k^2)],
    \label{eq:c00_zk}
\end{equation}
where $\omega_i^{(s)}$ are the relativistic omega integrals:\footnote{The following relation for $\omega_i^{(s)}$ is basically the same as \eqref{4.11}, just expressed in terms of more suitable integration variables.}
\begin{align}
    \omega_i^{(s)}(z_k)&=\frac{2\pi z_k^3}{K_2(z_k)^2}\int_0^{+\infty}d \psi\, \sinh^7\psi\, \cosh^i\psi\cdot\nonumber\\
    &\cdot K_j(2z_k\cosh\psi)\int_0^\pi d \theta\, \sin\theta\,\sigma_{kk}(\psi,\theta)(1-\cos^s\theta).
    \label{omega_i}
\end{align}
Here $\sigma_{kk}$ is the differential cross section for the interaction between two particles of the same species, and:
\begin{equation}
    \cosh\psi=\frac{\sqrt{s}}{2m}=\frac{\sqrt{(p_1+p_2)^2}}{2m_k},~~~j=\frac{5}{2}+\frac{1}{2}(-1)^i.
    \label{eq:cos_psi_e_j_omegaintegrals}
\end{equation}
Instead, in \eqref{1} and \eqref{2} the coefficients $\tilde{c}_{kl}$ and $\tilde{c}_{kk}$ denote contributions to the shear viscosity due to the interaction between different particle species:
\begin{align}
\tilde{c}_{12}=&\frac{32\rho^2c_1^2c_2^2}{3M_{12}^2n^2x_1x_2}[-10z_1z_2\zeta_{12}^{-1}Z_{12}^{-1}\omega_{1211}^{(1)}(\sigma_{12})-\nonumber\\
&10z_1z_2\zeta_{12}^{-1}Z_{12}^{-2}\omega_{1311}^{(1)}(\sigma_{12})+3\omega_{2100}^{(2)}(\sigma_{12})-\nonumber\\
    &3Z_{12}^{-1}\omega_{2200}^{(2)}(\sigma_{12})+Z_{12}^{-2}\omega_{2300}^{(2)}(\sigma_{12})],\nonumber\\
\tilde{c}_{11}=&\frac{32\rho^2c_1^2c_2^2}{3M_{12}^2n^2x_1x_2}[10z_1^2\zeta_{12}^{-1}Z_{12}^{-1}\omega_{1220}^{(1)}(\sigma_{12})+\nonumber\\
&10z_1^2\zeta_{12}^{-1}Z_{12}^{-2}\omega_{1320}^{(1)}(\sigma_{12})+3\omega_{2100}^{(2)}(\sigma_{12})-\nonumber\\
&3Z_{12}^{-1}\omega_{2200}^{(2)}(\sigma_{12})+Z_{12}^{-2}\omega_{2300}^{(2)}(\sigma_{12})],\nonumber\\
\tilde{c}_{22}=&\frac{32\rho^2c_1^2c_2^2}{3M_{12}^2n^2x_1x_2}[10z_2^2\zeta_{12}^{-1}Z_{12}^{-1}\omega_{1202}^{(1)}(\sigma_{12})+\nonumber\\
&10z_2^2\zeta_{12}^{-1}Z_{12}^{-2}\omega_{1302}^{(1)}(\sigma_{12})+3\omega_{2100}^{(2)}(\sigma_{12})-\nonumber\\
&3Z_{12}^{-1}\omega_{2200}^{(2)}(\sigma_{12})+Z_{12}^{-2}\omega_{2300}^{(2)}(\sigma_{12})].
\label{c_tilde_coefficients}
\end{align}

In the above relations $x_k=n_k/n$, where $n_k$ is the particle number density of particle type $k$, and $n=\sum_k n_k$ is the total particle number density. We have labeled by
\begin{align}
    M_{12}=m_1+m_2~~~~~\mu_{12}=m_1 m_2/M_{12}, \label{total_mass_and_reduced_mass}
\end{align}
the total mass and the reduced mass of the particles, respectively. Moreover,
\begin{align}
    \zeta_{12}=2\mu_{12}/T~~~~~Z_{12}=M_{12}/(2T). \label{Zeta_and_zeta}
\end{align}
The expression for the generalized relativistic omega integrals $\omega_{rtuv}^{(s)}$ is:
\begin{align}\omega_{rtuv}^{(s)}=&\frac{\pi \mu_{12}}{4T K_2(z_1)K_2(z_2)}\int_0^{+\infty}d \Psi_{12}\sinh^3\Psi_{12}\cdot\nonumber\\
&\cdot\left(\frac{g_{12}^2}{2\mu_{12}T}\right)^r\left(\frac{M_{12}}{P_{12}}\right)^t(\cosh \psi_1)^u (\cosh \psi_2)^v\cdot\nonumber \\
&\cdot K_\nu \left(\frac{P_{12}}{T}\right)\int_0^\pi d\theta \sin \theta \,\sigma_{12}(\Psi_{12},\theta)(1-\cos^s\theta),
\label{omega_rtuv}
\end{align}
where $\sigma_{12}$ is the differential cross section for the interaction among particles of the two species, and:
\begin{gather}
\Psi_{12}\equiv \psi_1+\psi_2,~~~\nu=\frac{5}{2}-\frac{1}{2}(-1)^{t+u+v},\nonumber\\
P_{12}^2=m_1^2+m_2^2+2m_1m_2\cosh\Psi_{12},\nonumber\\
g_{12}=\frac{m_1m_2\sinh\Psi_{12}}{P_{12}},\nonumber\\
\cosh\psi_1=\frac{1}{P_{12}}(m_1+m_2\cosh\Psi_{12}),\nonumber\\
\cosh\psi_2=\frac{1}{P_{12}}(m_2+m_1\cosh\Psi_{12})\label{Psi_Ptot_ecc}.
\end{gather}
Above $P_{12}=\sqrt{(p_1+p_2)^2}$ is the invariant center-of-mass energy of the two particles colliding with initial four momenta $p_1$ and $p_2$.\\

Unless otherwise stated (for instance in §\ref{Dependence on relative concentrations}), in our work the particle number concentrations $x_k$ and the relative mass concentrations $c_k$ have been obtained via a Boltzmann distribution:
\begin{align}
x_1(T)=&\frac{d_g \displaystyle\int \dd^3 \mathbf{p}\exp \left[ -{\sqrt{p^2+m_g^2}}/{T}\right] }{ \displaystyle \sum_{i=g,q} d_i\int \dd^3 \mathbf{p} \exp \left[ -{\sqrt{p^2+m_i^2}}/{T}\right]},\label{Boltzmann_concentrations}\\
x_2(T)=&1-x_1(T),\nonumber\\
c_1(T)=&\frac{m_g\, d_g \displaystyle \int \dd^3 \mathbf{p}\exp \left[ -{\sqrt{p^2+m_g^2}}/{T}\right]}{ \displaystyle \sum_{i=g,q} m_i \, d_i\int \dd^3 \mathbf{p} \exp \left[ -{\sqrt{p^2+m_i^2}}/{T}\right] },\nonumber\\
c_2(T)=&1-c_1(T).\nonumber
\end{align}
where $d_g=2\cdot(N_c^2-1)=16$ is the degeneration factor of gluons (spin $\times$ colour) and $d_q=2\cdot N_c \cdot  2 \cdot  N_f=36$ is the degeneration factor of quarks (spin $\times$ colour $\times$ antiparticles $\times$ flavours). It is worth noting that our reference to quarks and gluons is just exemplificatory, since the conclusions of the paper are much more general.\\

Let us note that, intuitively, if we consider a binary mixture in which both masses are equal $m_1=m_2=m$ and all the cross sections are equal as well $\sigma_{11}=\sigma_{22}=\sigma_{12}=\sigma$, what we get is an homogeneous gas, a single-component fluid whose particles have mass $m$ and interact with an interaction cross section $\sigma$. Of course, the CE formalism for a binary mixture has to reproduce this result in the aforementioned limits. For this purpose, in Appendix \ref{Appendix A} we perform the analytical calculations and show how \eqref{two_comp_eta} reduces to \eqref{4.9} when $m_1=m_2=m$ and $\sigma_{11}=\sigma_{22}=\sigma_{12}=\sigma$.

\begin{figure*}[ht]
\includegraphics[width=.9\textwidth]{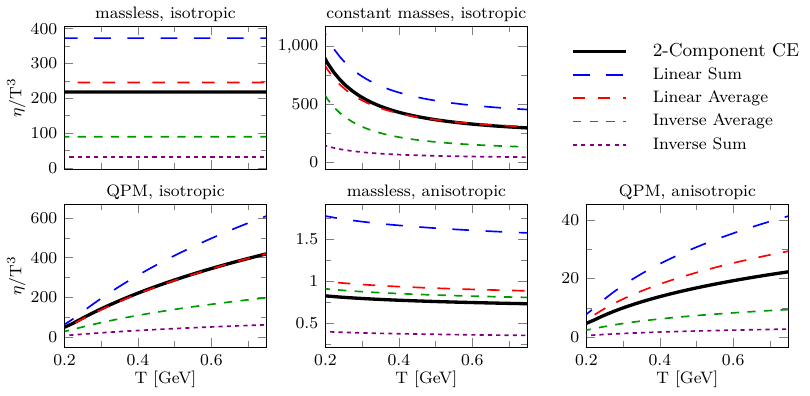}
\caption{Scaled shear viscosity $\eta/T^3$ vs $T$, for five different configurations of masses and cross sections. We compare the full 2-component result for $\eta$ (black full line) with various interpolations of the 1-component results for each of the two species. Both the five configurations and the various 1-component $\eta$ combinations are described in the main text, as well as in Appendices \ref{Appendix B} and \ref{Appendix E}.}
\label{eta_over_T3_vs_temperature}
\end{figure*}

\section{Comparison with single component interpolations}
\label{Results_averages}
As already discussed in the Introduction, we now study how well simple prescriptions involving only the single–component viscosities $\eta_1$ and $\eta_2$, given by Eq.\eqref{4.9}, can approximate the Chapman–Enskog results for a binary mixture. Such prescriptions correspond to four different rules that have been employed in the literature in different physical contexts.
In the plots that will follow, we will compare the full 2-component result with the following: 
\begin{itemize}
    \item Linear Sum, of the single component viscosities (blue):
        \begin{equation}
        \left.\eta\right|_{\text{L.S.}}=\eta_1+\eta_2.
            \label{eta_blue}
        \end{equation}
        This has been used, for instance, in \cite{Shternin:2008es}, where the shear viscosity of a mixture of electrons, muons and neutrons in a neutron star core is given by the sum of the respective viscosities, assuming small interaction between species. When the interaction between the species is not negligible, \eqref{eta_blue} is still used, but the intraspecies interaction is encoded in the respective relaxation times, as done also in \cite{Deng:2023rfw, Sarkar:2016gib}.
    \item Linear Average, i.e. by weighting the two viscosities with weights being given as in \eqref{Boltzmann_concentrations} (red):
        \begin{equation}
            \left.\eta\right|_{\text{L.A.}}=x_1\, \eta_1+x_2\, \eta_2.
            \label{eta_red}
        \end{equation}
    Such a relation has been applied, for instance, in the study of attractors in heavy-ion collisions, for a mixture of quarks and gluons \cite{Frasca:2024ege}. Check also \cite{Gorenstein:2007mw} for a use of this relation.
    
    \item Inverse Average, the inverse of the total viscosity is the weighted sum (with weights as in \eqref{Boltzmann_concentrations}) of the inverses of the single component viscosities (green):
\begin{equation}\left.\eta\right|_{\text{I.A.}}=\left(x_1\,\eta_1^{-1}+x_2\,\eta_2^{-1}\right)^{-1}.
            \label{eta_green}
        \end{equation}
        The use of this average is justified in the Relaxation Time Approximation (RTA) of the Boltzmann equation. Indeed, $\eta^i\sim \tau_{\text{relax}}^i$ for each component, and one can get an effective inverse relaxation time by summing the inverses of the $\tau_{\text{relax}}^i$ for all components, with respective weights.
    \item Inverse Sum, the inverse total viscosity is the sum of the inverses of the two viscosities (purple):
    \begin{equation} \left.\eta\right|_{\text{I.S.}}=\left(\eta_1^{-1}+\eta_2^{-1}\right)^{-1}.
        \label{eta_purple}
    \end{equation}
    Similarly as before, the total relaxation
frequency can be considered as the sum of the partial relaxation frequencies, as done by \cite{Manuel:2012rd} in the study of superfluid neutron stars. The same sort of formula has been used to describe the low $T$ behavior of the experimental values of the shear viscosity in the superfluid phase of a cold Fermi atomic gas in the unitarity limit, where one assumes that it is also dominated by the dynamics of the superfluid phonons \cite{Mannarelli:2012eg,Mannarelli:2012su}.

\end{itemize}
It is easy to prove that these relations always respect the following ordering:
\begin{equation}
\left.\eta\right|_{\text{L.S.}}\geq \left.\eta\right|_{\text{L.A.}} \geq 
\left.\eta\right|_{\text{I.A.}} \geq 
\left.\eta\right|_{\text{I.S.}}.
    \label{eq:ordering_of_interpolations}
\end{equation}
The middle inequality in \eqref{eq:ordering_of_interpolations} follows from the well-known Arithmetic Mean-Harmonic Mean inequality, and its equal sign holds if and only if $\eta_1=\eta_2$. The first and the third inequalities, instead, follow from the fact that the number concentrations trivially satisfy $x_1,x_2\leq 1$.

\subsection{Dependence on temperature}
\label{Dependence on temperature}

In Figure \ref{eta_over_T3_vs_temperature} we show the temperature dependence of the scaled shear viscosity $\eta/T^3$ of the binary mixture (in black), compared to the aforementioned single component combinations. We show the results for the following five different cases:
\begin{itemize}
    \item[-] \textit{massless, isotropic}: two component system with massless particles and isotropic cross sections (top-left panel of Fig. \ref{eta_over_T3_vs_temperature}); 
    \item[-]\textit{constant masses, isotropic}: two component system of massive particles with $m_1=0.8$ GeV, $m_2=0.5$ GeV and isotropic cross sections (top-middle panel of Fig. \ref{eta_over_T3_vs_temperature});
    \item[-] \textit{QPM, isotropic}: two component system with temperature dependent masses taken from the Quasi-Particle Model (QPM) and isotropic cross sections (bottom-left panel of Fig. \ref{eta_over_T3_vs_temperature}). The QPM is a microscopic effective model of QCD, which assumes a temperature dependence for the coupling constant and for the masses of quarks and gluons: more details on the QPM have been provided in Appendix \ref{Appendix B};
    \item[-] \textit{massless, anisotropic}: two component system with massless particles and angle-dependent cross sections (bottom-middle panel of Fig. \ref{eta_over_T3_vs_temperature}). The cross sections used are from tree-level pQCD;
    \item[-] \textit{QPM, anisotropic}: two component system with temperature dependent masses taken from the Quasi-Particle Model (QPM) and angle-dependent cross sections (bottom-right panel of Fig. \ref{eta_over_T3_vs_temperature}). The cross sections used are from tree-level pQCD applied to massive quarks and gluons, see Appendix \ref{Appendix D} for the technical details.
\end{itemize}

For a summary of the masses and the parameters hereby used, see Tab. \ref{table:paramters_fig1} in Appendix \ref{Appendix E}.

\begin{figure*}[t]
\includegraphics[width=.75\textwidth]{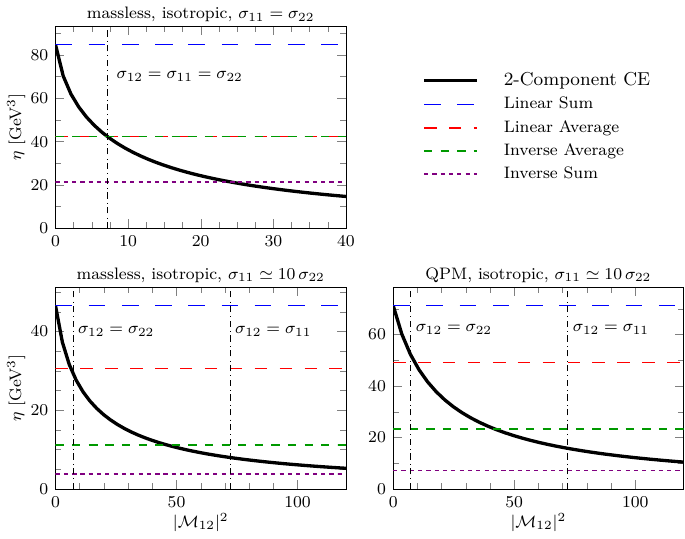}
\caption{Shear viscosity $\eta$ vs the squared scattering amplitude $|\mathcal{M}_{12}|^2$ of interaction between species 1 and 2, for three different configurations of masses and cross sections. We compare the full 2-component result for $\eta$ (black full line) with various interpolations of the 1-component results for each of the two species. Temperature is fixed at $T=0.5$ GeV. Both the three configurations and the various 1-component $\eta$ combinations are described in the main text, as well as in Appendices \ref{Appendix B} and \ref{Appendix E}.}
\label{eta_vs_M12squared}
\end{figure*}

The results for $\eta$ may vary significantly from one case to the other. It is important to stress, however, that here we want to focus on the relative position of the black curve with respect to all the others, rather than on the actual numerical values. What we observe is that in 4 out of 5 plots, the curve that better approximates the 2-component result is the red one, i.e. the Linear Average curve. In only one case (\textit{massless, anisotropic}) the green curve, representing the Inverse Average, is the best approximation. In any case, for all the plots we see that the 2-component result lies within or close to the band enclosed by the red and the green curves. We can interpret this as a consequence of the fact that the inter-species differential cross section $\sigma_{12}$, for all the cases considered, is of the same order of the cross sections within each species, namely $\sigma_{11}$ and $\sigma_{22}$. Apart from that, the quantitative agreement between the green, red and black curves may not be important, since it is not universal, rather is heavily dependent on the specific values of the cross sections.

The blue and the purple lines, as we will see in the next Subsections, are obtained only for extreme cases. For now it is sufficient to note that, as can be easily shown, for $\sigma_{12}\to 0$ Eq. \eqref{two_comp_eta} reduces to the Linear Sum $\eta\to\eta_1+\eta_2$, hence approaching the blue line from below. The opposite limit, $\sigma_{12}\to +\infty$ implies $\eta\to 0$. More details on such limit cases are explained in Appendix \ref{Appendix C}.

\subsection{Dependence on \texorpdfstring{$\sigma_{12}$}{sigma12}}
\label{Dependence on sigma12}
Due to the interesting behavior of the 2-component $\eta$ with respect to variations of $\sigma_{12}$, which we have briefly highlighted in §\ref{Dependence on temperature}, we now investigate further. In Figure \ref{eta_vs_M12squared} we plot the behavior of $\eta$ with respect to the squared scattering matrix $|\mathcal{M}_{12}|^2$, it being such that $|\mathcal{M}_{12}|^2/64 \pi^2 s=\sigma_{12}$. We show the results for three different cases:
\begin{itemize}
    \item[-] \textit{massless, isotropic}, $\sigma_{11}=\sigma_{22}$;
    \item[-] \textit{massless, isotropic}, $\sigma_{11}=(81/8)\,\sigma_{22}\simeq 10\, \sigma_{22}$;
    \item[-] \textit{QPM, isotropic}, $\sigma_{11}=(81/8)\,\sigma_{22}\simeq 10\, \sigma_{22}$.
\end{itemize}
The temperature has been fixed at $T=0.5$ GeV. For further details check once again Appendices \ref{Appendix B} and \ref{Appendix E}. As before, the 2-component CE approximation (black curve) is compared with various single component interpolations. Notice that in Fig. \ref{eta_vs_M12squared} the single component interpolations appear as flat lines, as expected since they do not depend on $\sigma_{12}$.

\begin{figure*}[ht]
\includegraphics[width=.85\textwidth]{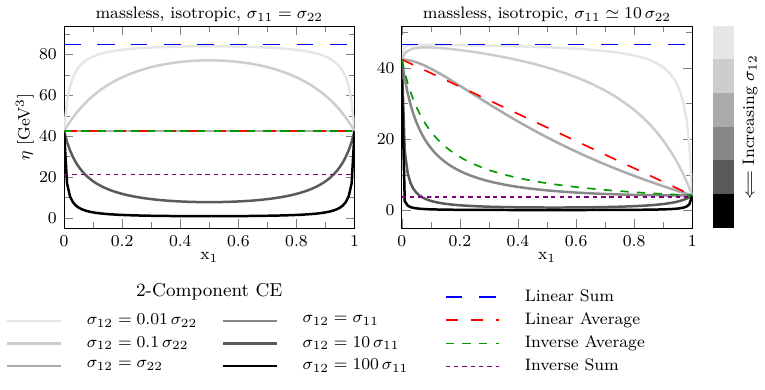}
\caption{Shear viscosity $\eta$ vs the number concentration $x_1$ of the species 1. We fix the cross sections $\sigma_{11}$ and $\sigma_{22}$, and show the full 2-component result for $\eta$ for different $\sigma_{12}$ (various shades of grey). Those 2-component results are compared with various interpolations of the 1-component results for each of the two species. Temperature is fixed at $T=0.5$ GeV. The details on the two configurations and on the various 1-component combinations are described in the main text and in Appendix \ref{Appendix E}.}
\label{eta_vs_concentrations}
\end{figure*}

To start, we see that for all cases considered, as already mentioned, for $\sigma_{12}\to 0$ the 2-component CE approaches the blue curve $\eta\to \left.\eta\right|_{\text{L.S.}}$. For increasing $|\mathcal{M}_{12}|^2$ the black curve decreases, and we find that for $\sigma_{12}$ between $\sigma_{22}$ and $\sigma_{11}$ the black curve lies around the region between the red and the green lines, i.e. between $\left.\eta\right|_{\text{L.A.}}$ and $\left.\eta\right|_{\text{I.A.}}$. This suggests that for small inter-species interaction (small $\sigma_{12}$) the Linear Average gives a good estimate for the $\eta$ of the mixture. In the massless case with isotropic $\sigma_{11}=\sigma_{22}$ the red and green lines are collapsed onto one, and the full result crosses those lines exactly when $\sigma_{12}=\sigma_{11}=\sigma_{22}$: this has to be expected, since when $\sigma_{11}=\sigma_{22}$ we have $\eta_1=\eta_2$, therefore green and red lines coincide since we are averaging equal quantities. Moreover, since in this particular case also $\sigma_{12}$ is equal to $\sigma_{11}$ and $\sigma_{22}$, our system is basically a single component gas, and therefore $\eta=\eta_1=\eta_2$ (check Appendix \ref{Appendix A}). For even greater $|\mathcal{M}_{12}|^2$, in particular for $\sigma_{12}$ greater than all the other scales in the system, the 2-component CE (black curve) gets closer to the Inverse Sum (purple), then it crosses this line and approaches zero in the limit $|\mathcal{M}_{12}|^2\to \infty$. It is important to emphasize, however, that the Inverse Sum is a good approximation only  for specific values of the inter-species interaction. So, it is a coincidence valid for some values of the parameters and does not imply a more general identity. As a final remark, notice that the considerations above hold for all three panels of Fig. \ref{eta_vs_M12squared}, regardless the mass of the interacting partons.

\subsection{Dependence on relative concentrations}
\label{Dependence on relative concentrations}
Up to now the relative concentrations of the two species have been fixed as in Eqs. \eqref{Boltzmann_concentrations}. We now want to relax this constraint and study the behavior of the shear viscosity with respect to the number concentrations. In Fig. \ref{eta_vs_concentrations} it is shown the behavior of $\eta$ with respect to the number concentration $x_1$ of the species 1. We show the results for two different cases:
\begin{itemize}
    \item[-] \textit{massless, isotropic}, $\sigma_{11}=\sigma_{22}$;
    \item[-] \textit{massless, isotropic}, $\sigma_{11}=(81/8)\,\sigma_{22}\simeq 10\, \sigma_{22}$.
\end{itemize}
The temperature has been fixed at $T=0.5$ GeV. For the actual values of the cross sections check Appendix \ref{Appendix E}. Here the full 2-component result is shown for different values of the inter-species cross section $\sigma_{12}$: from the top to the bottom of Figure \ref{eta_vs_concentrations}, as the shade of grey gets darker, $\sigma_{12}$ increases. As in the previous plots, those grey curves are compared with single component interpolations (since such interpolations do not depend on $\sigma_{12}$, the colored curves are unique). First of all, let us note that for increasing $\sigma_{12}$, the 2-component $\eta$ results decrease for every concentration $x_1$, except at the extremes $x_1=0$ and $x_1=1$: it is trivial to see that $\eta \to \eta_2$ for $x_1\to0$ and $\eta \to \eta_1$ for $x_1\to1$ whatever is $\sigma_{12}$, as expected. This suggests that the limits $\eta\to \left.\eta\right|_{\text{L.S.}}$ (i.e. blue line) for $\sigma_{12}\to 0$ and $\eta\to 0$ for $\sigma_{12}\to +\infty$, which we have previously mentioned, hold only for intermediate concentrations. In other terms, the sequence of functions $\eta_{\sigma_{12}}(x_1)$ for $\sigma_{12}\to 0$ or $\sigma_{12}\to+\infty$ converges pointwise but not uniformly in $x_1\in  ]0,1[$. As the previous results suggest, the behavior of $\eta$ in the limits $\sigma_{12}\to 0,+\infty$ and $x_1\to 0,1$ is quite tricky: in Appendix \ref{Appendix C} we elaborate further on these issues.

Figure \ref{eta_vs_concentrations} shows that, except at extreme concentrations $x_1$, for very small $\sigma_{12}$ the Linear Sum (blue curve) gives the best approximation. On the other hand, when $\sigma_{12}$ is very large with respect to $\sigma_{11}$ and $\sigma_{22}$, the best approximation is the Inverse Sum (purple curve). However, the Inverse Sum is the best for large $\sigma_{12}$ only because it gives the smallest $\eta$ among all kinds of formulas considered: it is the closest to zero, i.e. to the exact limit of the full two-component $\eta$ for $\sigma_{12}\to +\infty$ (black curve), and gives the best approximation only for this reason. Moreover, even if it appears to be the closest approximation, it is anyway very far in percentage from the CE results. As far as the ‘intermediate' $\sigma_{12}$ regime is concerned, we once again see that when $\sigma_{12}$ lies between $\sigma_{11}$ and $\sigma_{22}$, the corresponding $\eta$ is approximated with good accuracy by either the Linear Average (red curve), the Inverse Average (green curve) or none of them, depending on the specific values of the cross sections. In particular, by looking at Fig. \ref{eta_vs_concentrations} (right) we see that when $\sigma_{11}\simeq 10\,\sigma_{22}=10\,\sigma_{12}$, i.e. the third brightest shade of grey, the Linear Average (red curve) approximates quite well the 2-component result. Instead, when $\sigma_{12}=\sigma_{11}\simeq 10\,\sigma_{22}$ (fourth brightest shade of grey), the Inverse average (green curve) gives the best approximation. These results give useful insights on which interpolations work best for given regimes of cross sections. Finally, we also see that for some values of cross sections none of the single component interpolations gives a satisfactory approximation: see for instance the second brightest shade of grey in Fig. \ref{eta_vs_concentrations} (right), corresponding to $\sigma_{11}\simeq 10\, \sigma_{22}=10^2 \sigma_{12}$, which is not well approximated by any of the colored curves.

\section{Shear viscosity in Green-Kubo}
\label{sec:GK}
As for other transport coefficients, like heat-conductivity and bulk viscosity, also the shear viscosity $\eta$ can be related to the correlation function of the corresponding flux or tensor at thermal equilibrium \cite{Green:1954ubq,Kubo_1957}. In particular, this means that it is possible to derive these transport coefficients from the microscopic model, by using linear response theory. The underlying physical reason is that dissipation of fluctuations has the same physical origin as the relaxation towards equilibrium, therefore both dissipation and relaxation time are determined by the same transport coefficients \cite{Wiranata:2012br}.\\
In this framework, the expression of the shear viscosity is given by the Green-Kubo formula \cite{zubarev1996statistical}:
\begin{equation}
    \eta=\beta\lim_{T_{\text{max}}\to +\infty}\int_0^{T_{\text{max}}}\dd t \int_V \dd^3 \mathbf{x} \langle \pi^{xy}(\mathbf{x},t)\pi^{xy}(\mathbf{x},0)\rangle,
    \label{4.1}
\end{equation}
where $\beta$ is the inverse temperature and $\pi^{xy}$ is the $xy$ matrix element of the shear component of the energy momentum tensor. Here $\langle ... \rangle$ denote the following convolution procedure:
\begin{align}
\left\langle \pi^{xy}(\mathbf{x},t)\right.&\left.\pi^{xy}(\mathbf{x},0) \right\rangle =\nonumber \\
&\lim_{\mathcal{T}_{\text{max}} \to +\infty} \frac{1}{\mathcal{T}_{\text{max}}} \int_0^{\mathcal{T}_{\text{max}}} dt' \, \pi^{xy}(\mathbf{x},t + t')\pi^{xy}(\mathbf{x},t'). \label{eq:pi_pi_time_correlator}
\end{align}


Notice that the choice of the $xy$ component of $\pi$ is arbitrary: in \eqref{4.1}, as it can be shown, we can choose whatever indices $i,j=x,y,z$ as long as $i \neq j$. This is true since our system is isotropic, hence $\eta$ is a scalar.

In this paper, the shear stress correlations $\langle \pi^{xy}(\mathbf{x},t)\pi^{xy}(\mathbf{x},0)\rangle$ are computed through transport simulations of a particle system at thermal equilibrium. Such a system is confined in a static box of volume $V$ and subject to periodic boundary conditions, similarly to what has been done in \cite{Plumari:2012ep}. These simulations are performed employing a relativistic transport code developed, in recent years, to describe the Quark-Gluon Plasma (QGP) dynamics of heavy-ion collisions for different collision systems \cite{Ferini:2008he, Plumari:2012ep, Scardina:2012mik, Puglisi:2014sha, Scardina:2014gxa, Plumari:2015sia, Plumari:2015cfa, Scardina:2017ipo, Plumari:2019gwq, Plumari:2019hzp, Sun:2019gxg, Sambataro:2020pge, Sambataro:2022sns,Sambataro:2023tlv, Gabbana:2019uqv,Oliva:2020doe, Nugara:2023eku,Nugara:2024net}. In the most recent version the code has been written in \texttt{C} and has been optimized for High Performance Computing \cite{Nugara:2024net,Nugara:2023eku,Nugara:2025ueb}.
Our purpose is to numerically solve the relativistic Boltzmann transport equation with the full collision integral to evaluate the shear viscosity using the Green-Kubo formula, and to compare the results with the ones from the Chapman-Enskog formalism. 
In order to solve the transport equation we use the \textit{test particle method} \cite{Wong:1982zzb}, which basically consists in sampling the phase space distribution function using a large number of test particles.

Moving on to the computation of the shear component of the energy-momentum tensor, since there is no spatial inhomogeneity (e.g. external fields), we consider the volume-averaged shear tensor $\pi^{xy}(t)$. Its numerical evaluation is given by:
\begin{equation}
\pi^{xy}(t)=\,\frac{1}{V}\sum_{i=1}^{N}\frac{p_i^xp_i^y}{E_i},
    \label{eq:discretized_pi_xy}
\end{equation}
where $i$ runs over all the test particles. Once we do so, we finally derive the shear viscosity by using the following discretization of \eqref{4.1}:
\begin{equation}
\eta=\beta\,V \Delta t\sum_{j=0}^{N_{T_{\text{max}}}-1}\left\langle\pi^{xy}(j\Delta t) \pi^{xy}(0)\right\rangle,
    \label{4.1_discretized}
\end{equation}
where $N_{T_{\text{max}}}=T_{\text{max}}/\Delta t$, $T_{\text{max}}$ is the maximum time chosen in our simulation and the convolution procedure $\langle ...\rangle$ in \eqref{eq:pi_pi_time_correlator} has been performed as
\begin{align}
\left\langle \pi^{xy}(j\Delta t)\right.&\left.\pi^{xy}(0) \right\rangle=\nonumber\\
&\frac{1}{N_{\mathcal{T}_{\text{max}}}} \sum_{k=0}^{N_{\mathcal{T}_{\text{max}}}-1} \pi^{xy}(j\Delta t + k\Delta t)\pi^{xy}(k\Delta t)\label{eq:pi_pi_time_correlator_discretized}
\end{align}
where $N_{\mathcal{T}_{\text{max}}}=\mathcal{T}_{\text{max}}/\Delta t$ ($\Delta t$ is kept fixed for both \eqref{4.1_discretized} and \eqref{eq:pi_pi_time_correlator_discretized}). In order to have a good accuracy, $\mathcal{T}_{\text{max}}$ has to be chosen significantly larger than $T_{\text{max}}$. In particular, for each event we fix $\mathcal{T}_{\text{max}}/T_{\text{max}}=10$. The specific values of each quantity, however, have been fixed in order to ensure convergence of the result \eqref{4.1_discretized}. More details on the numerical implementation of the relativistic Boltzmann transport equation and of the Green-Kubo method can be found in \cite{Plumari:2012ep}.\\

\begin{figure}
    \centering
    \includegraphics[width=\columnwidth]{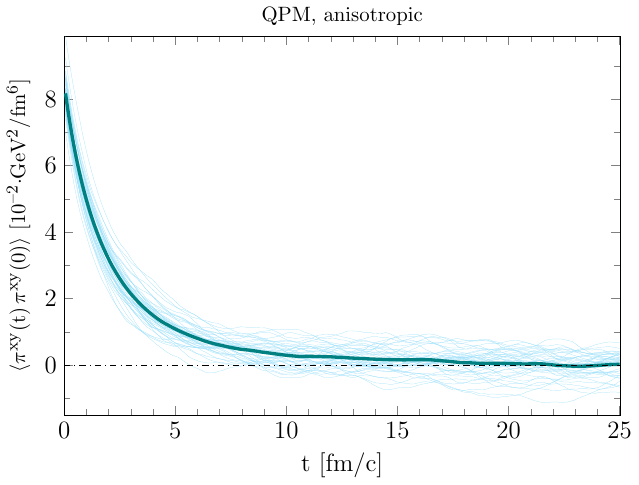}
    \caption{The behavior of $\langle \pi^{xy}(t) \pi^{xy}(0)\rangle $ with respect to $t$ at a temperature $T=0.5$ GeV in the \textit{QPM, anisotropic} case (check Appendices \ref{Appendix B} and \ref{Appendix E} for details). In cyan we show the results for 45 different numerical events, in dark blue their average, which highlights a clear exponential behavior.}
    \label{average_correlator_qpm_T_0.5}
\end{figure}

For illustrative purposes, in Figure \ref{average_correlator_qpm_T_0.5} we show (in cyan) the result of 45 different calculations for the time behavior of $\langle \pi^{xy}(t) \pi^{xy}(0)\rangle $ at a temperature of $T=0.5$ GeV in the \textit{QPM, anisotropic} case (check Appendices \ref{Appendix B} and \ref{Appendix E} for details). Together with the results for each event, which carry a significant amount of fluctuations, we show their ensemble average (dark blue), which highlights the typical exponential decay of such correlator.\footnote{The shear viscosity can also be obtained, as in \cite{Plumari:2012ep}, by assuming an exponential decay for the correlator, and then by linking the proper $\tau$ of such decay to $\eta$ using the Green-Kubo formula \eqref{4.1}.} Notice that the negative values of the correlators are only a matter of numerical fluctuations. Indeed, they basically disappear in the averaged correlator and would be suppressed by further increasing the number of test particles.

Let us report the parameters we will use in Section \ref{Results for etaovers} in the Green-Kubo simulation. For each value of the temperature, the Green-Kubo values for $\eta$ have been obtained as an average over 45 numerical events. Each event would run with a total number of test particles of about $1.5 \cdot 10^6$. The value of the grid parameter used in this work is $\Delta x=\Delta y=\Delta z=0.4 $ fm, with a total box volume of $V=(5.2$ fm)$^3$. The timestep has been fixed as $\Delta t=0.0625$ fm/c, whereas the total time $T_{\text{max}}$ has been varied for each value of the temperature: indeed, $T_{\text{max}}$ has to be fixed large enough so that the exponential behavior of each correlator is appropriately dumped, but not too large in order to exclude the noise which arises from the statistical fluctuations of each event (those fluctuations have a major impact when the correlator approaches its vanishing value at late times). Since the timescales of such behaviors change from one temperature value to the other, this required varying $T_{\text{max}}$.

\section{Comparison between Chapman-Enskog and Green-Kubo}
\label{Results for etaovers}
Once we highlighted the details of the Green-Kubo procedure for the evaluation of the shear viscosity, we want to use this method as a benchmark for our Chapman-Enskog calculations for a binary mixture. By keeping in mind applications to heavy-ion collisions physics, we want to compare the temperature dependence of $\eta/s$, being $s$ the entropy density, with Lattice QCD results \cite{Meyer:2007ic, Meyer:2009jp, Mages:2015rea, Borsanyi:2018srz, Altenkort:2022yhb}.

More specifically, for both Green-Kubo and Chapman-Enskog calculations we performed the calculations in the \textit{QPM, anisotropic} case (see Appendix \ref{Appendix E}). In the transport code we simulate the evolution of $u, \Bar{u}, d,\Bar{d}, s, \Bar{s}$ quarks, each weighted as $2\cdot N_c=6$ (spin $\times$ colour), and gluons, each weighted as $2\cdot (N_c^2-1)=16$ (spin $\times$ colour). The interaction matrices depend on the species colliding, since we may or may not have the contribution of annihilation/exchange diagrams. The full scattering matrices are derived in Appendix \ref{Appendix D}. The entropy density $s$ is evaluated in the QPM as
\begin{equation}
    s=\frac{\varepsilon_{\text{QPM}}+P_{\text{QPM}}}{T},
    \label{entropy_density_QPM}
\end{equation}
$\varepsilon_{\text{QPM}}$ and $P_{\text{QPM}}$ being given by equations \eqref{eps_QPM} and \eqref{P_QPM}, respectively. Once again, check Appendix \ref{Appendix B} for details.\\

\begin{figure}[t]
    \centering
\includegraphics[width=\columnwidth]
{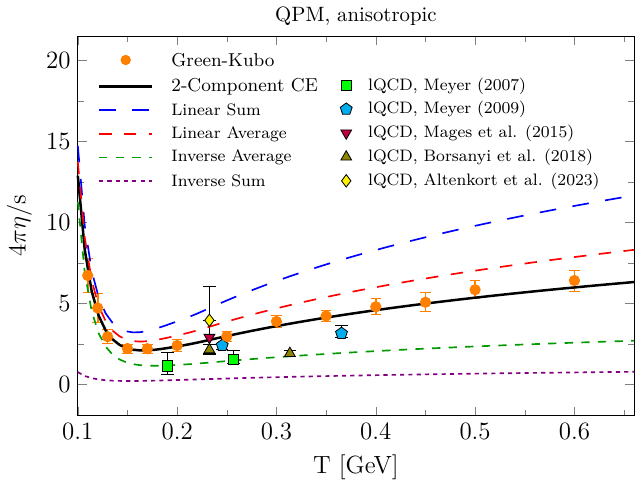}
    \caption{The ratio shear viscosity over entropy density $4\pi \eta/s$ in the full 2-component Chapman-Enskog approximation (black full line), compared to the Green-Kubo results, to various combinations of the 1 component results for each of the two species, and various lattice QCD data from \cite{Meyer:2007ic} (green squares), \cite{Meyer:2009jp} (cyan pentagon), \cite{Mages:2015rea} (purple downward triangle), \cite{Borsanyi:2018srz} (brown upward triangles) and \cite{Altenkort:2022yhb} (yellow losange).
    The calculations have been performed in the \textit{QPM, anisotropic} case, all the parameters are listed in Table \ref{table:paramters_fig5} (check Appendices \ref{Appendix B} and \ref{Appendix E} for further details).}
    \label{eta/s_averaged}
\end{figure}

The results are shown in Figure \ref{eta/s_averaged}. Along with the full 2-component CE approximation and the Green-Kubo results, we show different combinations of the 1-component viscosities of the two species, and various lattice QCD data from \cite{Meyer:2007ic, Meyer:2009jp, Mages:2015rea, Borsanyi:2018srz, Altenkort:2022yhb}. The comparison between Chapman-Enskog (black solid line) and Green-Kubo results (orange points) is quite good within the error bars. This shows that 1$^\text{st}$ order CE is already a quite good approximation. A 2$^\text{nd}$ order Chapman-Enskog approximation for a binary mixture is certainly doable in principle, even though that would require much more cumbersome calculations. In any case, the agreement in Figure \ref{eta/s_averaged} suggests that higher order CE approximations may not significantly modify the 1$^\text{st}$ order result.

Moving on to the single component interpolations, we see that Linear Sum (dashed blue line), Inverse Sum (dashed purple line) and Inverse Average (dashed green line) show large deviations with respect to Green-Kubo in the whole range of temperatures explored. Among the four single component interpolations, the Linear Average result (dashed red line) is the one showing the best agreement, both qualitatively and quantitatively, with the orange points. In light of the discussions of the previous Sections, this result suggests that in the physical context of QGP (which is the one that the \textit{QPM, anisotropic} case aims to reproduce) the quark-quark, quark-gluon and gluon-gluon cross sections have similar relevance in the estimation of $\eta$ among the different channels. In any case, what we see is that only the full 2-component Chapman-Enskog calculation (black solid line) is able to accurately reproduce the Green-Kubo results.

Even though this is not the main focus of the present work, we also note that the full CE curve is consistent with the various lattice data, within the large uncertainties. The minimum that we find for the $\eta/s$ (i.e. $4\pi \eta/s\sim2$ for $T\sim 0.17$ GeV) is higher with respect to the conjectured minimum value of
$4\pi \eta/s=1$ from AdS/CFT \cite{Kovtun:2004de}. On the other hand, our minimum is in agreement with hydrodinamical/transport calculations \cite{Noronha-Hostler:2015dbi, Roch:2020zdl, Plumari:2015cfa, Ozvenchuk:2012kh}: when applied to experimental data of collective flows at RHIC and LHC, these calculations suggest a value of $4\pi \eta/s\sim 2$.

\section{Conclusions and Outlook}
\label{sec:concl}
In this work, we have investigated the shear viscosity of a binary mixture within a relativistic kinetic theory. The theoretical framework is based on the 1$^\text{st}$ order Chapman–Enskog expansion, which allows for a systematic treatment of transport coefficients for binary systems. In particular, we compare the $\eta$ of a binary mixture against various combinations of the viscosities of the single component Chapman-Enskog results, $\eta_1$ and $\eta_2$. 

The general picture is that 
each of the single component approximations can reproduce the full 2-component results only within specific ranges of the parameters (masses, cross sections, concentrations). Across the various cases under study we have seen that, when all the cross sections are similar within a factor of 2, the $\eta$ of the 2-component CE is well approximated by the linear weighted average $x_1\, \eta_1+x_2\,\eta_2$ of the single component viscosities (here $x_1$ and $x_2$ are the number concentrations of each component, such that $x_1+x_2=1$). Another good approximation is provided by the weighted average of the inverses, i.e. $1/\left(x_1\,\eta_1^{-1}+x_2\,\eta_2^{-1}\right)$: in a specific case (i.e. the \textit{massless, anisotropic} case in Fig. \ref{eta_over_T3_vs_temperature}) it turned out to be even slightly better than the linear weighted average. To better understand these results, we investigated the behavior of our quantities by manually tuning the number concentrations $x_1,x_2$ and the interspecies cross section $\sigma_{12}$. What we found is that the linear and the inverse weighted averages are good approximations just because $\sigma_{12}$ is of the same order of the interaction cross sections among each species, i.e. $\sigma_{12}\sim \sigma_{11}$ and $\sigma_{12}\sim\sigma_{22}$. In the extreme case $\sigma_{12}\to 0$, we instead see that $\eta$ of the mixture converges exactly to the simple sum $\eta=\eta_1+\eta_2$.
Overall, our study allows to understand which single component approximation works the best in reproducing the full 2-component result, for any given range of parameters. Our formalism also shows that in some regimes none of the single component approximations provides a satisfactory approximation of the 2-component result.

Once we have made such comparisons, we have checked whether the 2-component result or any of the 1-component averages would be able to properly describe the viscosity over entropy density ratio $\eta/s$ in a physical case. By developing an independent numerical calculation using Boltzmann transport equation, we used the Green-Kubo formula to have another estimate of the shear viscosity, as an external cross-check. When the $\eta/s$ of the QGP is computed within a quasi-particle model, we see that the $\eta/s$ cannot be trivially described by any combination of the single species (quark and gluon) $\eta/s$, because the quark-quark, quark-gluon and gluon-gluon interaction cross sections are of comparable relevance. Among all the formulas considered here, we see that the Linear Average is the single-component combination which better approximates Green-Kubo points. However, only the full 2-component Chapman-Enskog formula accurately reproduces the Green-Kubo results, within the error bars, for all temperatures considered.\\

This work sheds light on the various approximations used to evaluate the shear viscosity of a mixture, when one does not have sufficient information on the interaction between one species and another. In particular, we have shown that the conjectured formulas that try to estimate $\eta$ as combinations of the single fluid components can largely underestimate or overestimate the ‘exact' result of the 2-component fluid. The purpose of the study is general, since its conclusions are not limited to hot QCD matter, but may be applied also to the study of neutron stars and condensed matter systems. In the study of neutron stars, Pauli blocking effects are quite substantial \cite{Typel:2009sy}, therefore a future extension of this study to include these effects has to take into account a Boltzmann equation with the inclusion of the Uehling–Uhlenbeck terms in the collision integral \cite{Wang:2023gta}. In light of quark-gluon plasma applications, it may be useful to explore the multi-component ($\geq3$) Chapman-Enskog formalism, in order to distinguish the strange flavor from the light flavors and the gluons: the outcomes may then once again be compared to the Green-Kubo results. 

\subsection*{Acknowledgments}
The authors thank A. Beraudo, S. Burrello, F. Frascà, M. Mannarelli, L. Oliva and M. Prakash for help and useful discussions. We acknowledge the funding from UniCT under PIACERI ‘Linea di intervento 1’ (M@uRHIC). This work has been partly funded by 2022SM5YAS ‘‘Advanced probes of the Quark Gluon Plasma'', under the program PRIN 2022 PNRR of the Italian Ministero dell’Università e Ricerca (MUR).

\onecolumngrid

\appendix

\section{Analytical calculations for the 2-component to 1-component reduction of \texorpdfstring{$\eta$}{eta}}
\label{Appendix A}
As already mentioned in §\ref{binary mixture}, if we consider a 2-component mixture in which $m_1=m_2\equiv m$ and $\sigma_{12}=\sigma_{11}=\sigma_{22}\equiv \sigma$, what we obtain is trivially a single component fluid, whose particles have mass $m$ and interact with a differential cross section $\sigma(\psi,\theta)$. It is therefore natural to expect that if we consider our 2-component formula \eqref{two_comp_eta} for $\eta$ and put $m_1=m_2=m$ and $\sigma_{12}=\sigma_{11}=\sigma_{22}\equiv \sigma$, we find the 1-component result \eqref{4.9}, independently on the relative concentrations $x_1,x_2$. In this Appendix we analytically show that the Chapman-Enskog formalism outlined in §\ref{binary mixture} recovers this result.\\

Let us preliminary study the relation between the generalized two component omega integrals \eqref{omega_rtuv} and the single component omega integrals \eqref{omega_i} (which is the same as \eqref{4.11}) when all the masses and cross sections are the same.
For $m_1=m_2=m$ we have $\mu_{12}=m_1m_2/(M_{12})=m/2$ by the virtue of \eqref{total_mass_and_reduced_mass}. By taking into account the relations \eqref{Psi_Ptot_ecc} we can write \eqref{omega_rtuv} as:
\begin{align}\omega_{rtuv}^{(s)}=&\frac{\pi \cdot m/2}{4T [K_2(z)]^2}\int_0^{+\infty}d \Psi \sinh^3\Psi\left[\frac{m^4\sinh^2\Psi}{2(m/2)T P^2}\right]^r\left(\frac{M}{P}\right)^t\cdot\nonumber \\
&\left[\frac{1}{P}m(1+\cosh \Psi)\right]^u \left[\frac{1}{P}m(1+\cosh \Psi)\right]^v K_\nu \left(\frac{P}{T}\right)\int_0^\pi d\theta \sin \theta \,\sigma(\Psi,\theta)(1-\cos^s\theta).
\label{omega_rtuv_A}
\end{align}
By considering that $\Psi=2\psi$ in \eqref{Psi_Ptot_ecc}, we can rewrite:
\begin{align}
    \sinh\Psi&=2\sinh\psi\cosh\psi,\nonumber\\
    P&=\sqrt{2m^2(1+\cosh\Psi)}=\sqrt{2m^2(2\cosh^2\psi)}=2m\cosh\psi,\nonumber\\
    \frac{M}{P}&=(\cosh\psi)^{-1},\nonumber\\
    \frac{1}{P}m(1+\cosh \Psi)&=\cosh\psi,
    \label{from_Psi_to_psi}
\end{align}
and \eqref{omega_rtuv_A} then takes the form:
\begin{align}\omega_{rtuv}^{(s)}=&\frac{\pi m}{8T [K_2(z)]^2}\int_0^{+\infty}(2\,d \psi) (\cosh^2\Psi-1)(2\sinh\psi\cosh\psi)\left(\frac{m}{T}\right)^r \cdot\nonumber\\
&(\sinh^2\psi)^r (\cosh\psi)^{-t+u+v} K_\nu \left(2z\cosh\psi\right)\int_0^\pi d\theta \sin \theta \,\sigma(\psi,\theta)(1-\cos^s\theta).
\label{omega_rtuv_B}
\end{align}
Moreover, by noting that $\cosh^2\Psi-1=4\cosh^2\psi\sinh^2\psi$, we finally obtain:
\begin{align}\omega_{rtuv}^{(s)}=\frac{2\pi z^{1+r}}{[K_2(z)]^2}\int_0^{+\infty}d\psi\, (\sinh\psi)^{3+2r} (\cosh\psi)^{3-t+u+v} K_\nu \left(2z\cosh\psi\right)\int_0^\pi d\theta \sin \theta \,\sigma(\psi,\theta)(1-\cos^s\theta).
\label{omega_rtuv_C}
\end{align}
A quick comparison with \eqref{omega_i} finally shows the relation between the two component $\omega_{rtuv}^{(s)}$ and the single component $\omega_{i}^{(s)}$ when masses and cross sections are equal:
\begin{equation}
\omega_{2,t,u,v}^{(s)}=\omega_{3-t+u+v}^{(s)}.
    \label{Omega_to_omega}
\end{equation}
Using this preliminary calculation, we can now see what happens to the coefficients $\tilde{c}$ in \eqref{c_tilde_coefficients}, of which we report the general expression, just for convenience:
\begin{align}
\tilde{c}_{12}=\frac{32\rho^2c_1^2c_2^2}{3M_{12}^2n^2x_1x_2}&[-10z_1z_2\zeta_{12}^{-1}Z_{12}^{-1}\omega_{1211}^{(1)}(\sigma_{12})-10z_1z_2\zeta_{12}^{-1}Z_{12}^{-2}\omega_{1311}^{(1)}(\sigma_{12})+\nonumber\\
&3\omega_{2100}^{(2)}(\sigma_{12})-3Z_{12}^{-1}\omega_{2200}^{(2)}(\sigma_{12})+Z_{12}^{-2}\omega_{2300}^{(2)}(\sigma_{12})],\nonumber\\
\tilde{c}_{11}=\frac{32\rho^2c_1^2c_2^2}{3M_{12}^2n^2x_1x_2}&[10z_1^2\zeta_{12}^{-1}Z_{12}^{-1}\omega_{1220}^{(1)}(\sigma_{12})+10z_1^2\zeta_{12}^{-1}Z_{12}^{-2}\omega_{1320}^{(1)}(\sigma_{12})+\nonumber\\
&3\omega_{2100}^{(2)}(\sigma_{12})-3Z_{12}^{-1}\omega_{2200}^{(2)}(\sigma_{12})+Z_{12}^{-2}\omega_{2300}^{(2)}(\sigma_{12})],\nonumber\\
\tilde{c}_{22}=\frac{32\rho^2c_1^2c_2^2}{3M_{12}^2n^2x_1x_2}&[10z_2^2\zeta_{12}^{-1}Z_{12}^{-1}\omega_{1202}^{(1)}(\sigma_{12})+10z_2^2\zeta_{12}^{-1}Z_{12}^{-2}\omega_{1302}^{(1)}(\sigma_{12})+\nonumber\\
&3\omega_{2100}^{(2)}(\sigma_{12})-3Z_{12}^{-1}\omega_{2200}^{(2)}(\sigma_{12})+Z_{12}^{-2}\omega_{2300}^{(2)}(\sigma_{12})].
\label{c_tilde_coefficients_appendix}
\end{align}
By looking at the definition of the 2-component omega integrals (the original definition in \eqref{omega_rtuv} or equivalently the simplified version \eqref{omega_rtuv_A}), it can be immediately seen that when the masses are equal we have:
\begin{equation}
\omega_{1211}^{(1)}=\omega_{1220}^{(1)}=\omega_{1202}^{(1)},~~~~~~~~\omega_{1311}^{(1)}=\omega_{1320}^{(1)}=\omega_{1302}^{(1)}.
    \label{omega_equalities}
\end{equation}
Similarly, by using the relation \eqref{Omega_to_omega} we have just derived, we have
\begin{equation}
\omega_{2100}^{(2)}=\omega_{2}^{(2)},~~~~~~~~\omega_{2200}^{(2)}=\omega_{1}^{(2)},~~~~~~~~\omega_{2300}^{(2)}=\omega_{0}^{(2)}.
    \label{Omega_to_omega_application}
\end{equation}
By using \eqref{omega_equalities} and \eqref{Omega_to_omega_application}, the equations \eqref{c_tilde_coefficients_appendix} can be written as:
\begin{align}
\tilde{c}_{12}=&\frac83 x_1 x_2[-10 \omega_{1211}^{(1)}-10 z^{-1}\omega_{1311}^{(1)}+3\omega_{2}^{(2)}-3z^{-1}\omega_{1}^{(2)}+z^{-2}\omega_{0}^{(2)}]=\nonumber\\
=&\frac83 x_1 x_2[-10A+3B],\nonumber\\
\tilde{c}_{11}=\tilde{c}_{22}=&\frac83 x_1 x_2[10 \omega_{1211}^{(1)}+10 z^{-1}\omega_{1311}^{(1)}+3\omega_{2}^{(2)}-3z^{-1}\omega_{1}^{(2)}+z^{-2}\omega_{0}^{(2)}]=\nonumber\\
=&\frac83 x_1 x_2[10A+3B],\label{c_tilde_coefficients_appendix_2}
\end{align}
having defined $A\equiv \omega_{1211}^{(1)}+z^{-1}\omega_{1311}^{(1)}$ and $B\equiv\omega_{2}^{(2)}-z^{-1}\omega_{1}^{(2)}+z^{-2}\omega_{0}^{(2)}/3$.\\

Now we can tackle the full expression \eqref{two_comp_eta}: we will simplify the numerator and the denominator separately. Starting from the numerator we have:
\begin{align}
\gamma_1^2c_{22}+\gamma_2^2c_{11}-2\gamma_1\gamma_2c_{12}=&\left(10\frac{K_3(z)}{K_2(z)}\right)^2\left[x_1^2 c_{22}+x_2^2 c_{11}-2x_1x_2 c_{12}\right]=\nonumber\\
=&\left(10\frac{K_3(z)}{K_2(z)}\right)^2x_1x_2\left[16x_1x_2B+\frac{80}{3} x_1^2 A+8 x_1^2 B+16x_1x_2B+\right.\nonumber\\
&\left.\frac{80}{3} x_2^2 A+8 x_2^2B+\frac{160}{3}x_1x_2A-16 x_1x_2B\right]=\nonumber\\
=&\left(10\frac{K_3(z)}{K_2(z)}\right)^2x_1 x_2\left[\frac{80}{3}A+8B\right].
    \label{numerator_of_2comp_eta}
\end{align}
Moving on to the denominator we instead have:
\begin{align}
c_{11}c_{22}-c_{12}^2= & \left[16x_1^2B+\frac83 x_1 x_2(10A+3B)\right]\left[16x_2^2B+\frac83 x_1 x_2(10A+3B)\right]-
\left[\frac83 x_1 x_2(-10A+3B)\right]^2 =\nonumber\\
= & 256x_1^2 x_2^2 B^2+16B \frac{8}{3}x_1x_2(10A+3B)(x_1^2+x_2^2)+\frac{64}{9}x_1^2 x_2^2\left[(10A+3B)^2-(-10A+3B)^2\right]=\nonumber\\
= & 256x_1^2 x_2^2 B^2+\frac{64}{9}x_1x_260AB(x_1^2+x_2^2)+128x_1x_2B^2(x_1^2+x_2^2)+\frac{64}{9}x_1^2 x_2^2 120AB=\nonumber\\
= & \frac{64}{9}x_1x_260AB+128x_1x_2B^2=\nonumber\\
= & 16x_1 x_2B\left[\frac{80}{3}A+8B\right].
    \label{denominator_of_2comp_eta}
\end{align}
This means that, by using the results \eqref{numerator_of_2comp_eta} and \eqref{denominator_of_2comp_eta}, the equation \eqref{two_comp_eta} becomes:
\begin{equation}
    \eta=\frac{T}{10}\frac{\gamma_1^2c_{22}+\gamma_2^2c_{11}-2\gamma_1\gamma_2c_{12}}{c_{11}c_{22}-c_{12}^2}=\frac{T}{10}\frac{\left[10K_3(z)/K_2(z)\right]^2x_1 x_2\left[80A/3+8B\right]}{16x_1 x_2B\left[80A/3+8B\right]}=\frac{T}{10} \frac{\left[10K_3(z)/K_2(z)\right]^2}{16B}=\frac{T}{10}\frac{\gamma_0^2}{c_{00}},
\end{equation}
i.e. exactly equation \eqref{4.9}. This means that, when the masses $m_1,m_2$ of the two components are equal and all the differential cross sections $\sigma_{11}, \sigma_{22}$ (within each species) and $\sigma_{12}$ (interaction between species) are equal as well, the 2-component result reduces exactly to the 1-component formula. As expected, this occurs independently on the relative concentrations $x_1,x_2$, since in this limit there is nothing differentiating the two components.

\section{Details on the Quasi Particle Model}
\label{Appendix B}
It is well known that perturbative QCD is applicable only for energies greater than the typical scale $\Lambda_{\text{QCD}}$. However, the temperature range achieved at RHIC and LHC is not sufficiently large compared to the $\Lambda_{\text{QCD}}$, which implies that the coupling $\alpha_s$ is not sufficiently small and we cannot overlook the non-perturbative aspects of the interaction. Consequently, the medium formed in the collision can be described as a many-body system which is heavily screened and influenced by non-perturbative effects, even at very high temperatures. As well known, the experimental data related to heavy quark dynamics, such as the $D$ meson nuclear modification factor $R_{AA}$ and elliptic flow $v_2$, are not well reproduced by perturbative calculations. In particular, perturbative approaches effectively describe phenomena occurring at high transverse momenta $p_T$, but underestimate interactions at low $p_T$ (below 15-20 GeV). In order to include non-perturbative aspects of the deconfined state, one should therefore move beyond the notion of massless and weakly interacting bulk partons.

One approach proposed consists in describing quarks and gluons in the Quark-Gluon Plasma as an ensemble of \textit{quasi-particles} near the critical temperature. This approach allows to deal with the appropriate degrees of freedom while still describing the lattice QCD results of equilibrium thermodynamics (e.g. the EoS). By attributing most interactions to the $T$-dependent effective masses, we can treat the remaining interaction among quasi-particles perturbatively. As shown in recent years, the inclusion of the quasi-particle model in relativistic heavy-ion simulations, along with the incorporation of an appropriate hadronization scheme for coalescence and fragmentation, has led, among other things, to a better description of the main observables of heavy hadrons and a more consistent estimation of the spatial diffusion coefficient with values extracted from lattice QCD \cite{Plumari:2017ntm}.\\

More specifically, within the Quasi Particle Model (QPM), we assume the masses for gluons and light quarks to have the following temperature dependence:
\begin{equation}
m_g(T)^2=\frac{1}{6}\left(N_c+\frac12 N_f\right)g(T)^2T^2,~~~~~m_{q}(T)^2=\frac{N_c^2-1}{8N_c}g(T)^2T^2,
\label{3.9}
\end{equation}
where $N_c=3$ denotes the number of colors, $N_f=3$ is the number of flavors and we assume no chemical potential. The function $g(T)$ is an effective coupling constant, whose temperature dependence will be derived by fitting lattice QCD data.

In this model the total pressure of the system can be written as the sum of independent contributions from the different constituents (each having the $T$-dependent effective mass in \eqref{3.9}). However, when the temperature-dependent masses are included in the pressure, its derivative with respect to temperature will produce an extra term in the energy density, which does not have the ideal gas form. Therefore, the model is completed by introducing a temperature-dependent ‘bag' term $B(T)$ to account for further non-perturbative effects and to ensure thermodynamic consistency \cite{Plumari:2011mk}: 
\begin{align}
\varepsilon_{\text{QPM}}(T)&=\sum_id_i \int \frac{\dd^3 \mathbf{p}}{(2\pi)^3}E_i(p) f_i(p)+B(T)
    \label{eps_QPM},\\
    P_{\text{QPM}}(T)&=\sum_{i}d_i\int \frac{\dd^3 \mathbf{p}}{(2\pi)^3}\frac{p^2}{3E_i(p)}f_i(p)-B(T).\label{P_QPM}
\end{align}
In the above equations $f_i(p)=[1\mp \exp[\beta E_i(p)]]^{-1}$ are the Bose/Fermi distribution functions for gluons and quarks, respectively, $E_i(p)=\sqrt{\mathbf{p}^2+m_i^2}$ and $d_i$ are the degeneracy factors, equal to $2\cdot 2\cdot N_c$ for quarks and to $2\cdot (N_C^2-1)$ for gluons. Each of the above quantities has been expressed in terms of two unknown functions, namely $g(T)$ and $B(T)$. However, only one of those functions is actually independent, since we have to enforce thermodynamic consistency by imposing:
\begin{equation}
\left.\frac{\partial P_{\text{QPM}}}{\partial m_i}\right|_{T}=0,~~~~ i=u,d,s,g.
    \label{eq:thermodynamic_constistency}
\end{equation}
At this point, only $g(T)$ has to be determined, and this is done by performing a fit to the lattice data from \cite{Borsanyi:2010cj}:
\begin{equation}
    \varepsilon_{\text{QPM}}(T)=\varepsilon_{\text{lattice}}(T),
    \label{3.11}
\end{equation}

\begin{figure}[t]

\includegraphics[width=0.65\textwidth]{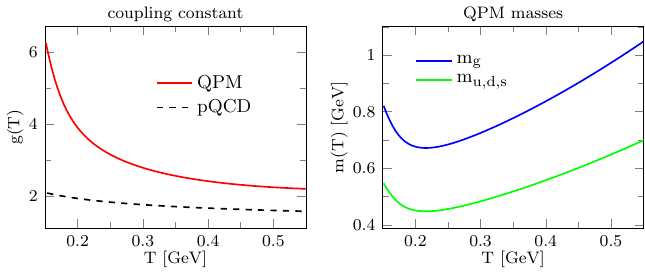}

\caption{Left panel: the coupling $g$ as a function of $T$ for pQCD calculation (black dashed line, from \cite{Kaczmarek_2005}) and QPM (red solid line, for $N_f=3$). Right panel: the gluon and quark Quasi-Particle masses as functions of $T$.}
\label{fig1.1}
\end{figure}
By doing so we get the $g(T)$ in the left panel of Figure \ref{fig1.1}: as shown, the QPM recovers the correct pQCD limit for high temperature. For $T_0=0.151$ GeV and $g_0=6.26$ (the initial value for $g_0$ has been set in order to match the lattice data for the pressure at temperature $T_0$) this function can be fitted as:
\begin{align}
    g(T)&=(c_0+c_1T+c_2T^2)\exp[-c_3\, T^{c_4}]~~~~\,\text{for }T\leq 0.22 \text{ GeV},\nonumber\\
g(T)&=\sqrt{\frac{16\pi^2}{9\log[(b_0(T/b_1-b_2))^2]}}~~~~~~~~~\,\,\,\,\text{ for }T> 0.22 \text{ GeV},
\label{eq:g_fit}
\end{align}
where the values of the constants are reported in Table \ref{tab:1}, assuming that $T$ is considered in GeV.
\begin{table}[H]
    \centering
    \begin{tabular}{|c|c|c|c|c|c|c|c|}
        \hline
         $c_0$&$c_1$&$c_2$&$c_3$&$c_4$&$b_0$&$b_1$&$b_2$  \\\hline
$46.737$&$-436.86$&$1266.4$&$77.763$&$2.7523$&$2.2800$&$0.155$&$0.55$ \\\hline
    \end{tabular}
    \caption{Values for the constants used in the fit for $g(T)$.}
    \label{tab:1}
\end{table}

In the right panel of Figure \ref{fig1.1} we instead show the quark and gluon masses in \eqref{3.9} that we obtain from this procedure. We notice that at sufficiently high temperatures $m(T)\sim T$, as one can expect since in these regimes $T$ remains the only scale of the system.

More details on the QPM, as well as on its possible extension to a momentum dependence in the masses (known as QPMp), can be found in \cite{Sambataro:2024mkr}.

\section{Shear viscosity in the limit case \texorpdfstring{$\sigma_{12}\to 0$}{sigma to zero}}
\label{Appendix C}
In Section \ref{Results_averages} we have highlighted more than once that, when $\sigma_{12}$ gets smaller, the results for the shear viscosity of a binary mixture approach the blue curves, i.e. the simple sum of the two viscosities (see Figures \ref{eta_vs_M12squared} and \ref{eta_vs_concentrations}). This is very easy to check also analytically, since when $\sigma_{12}\to 0$ Eq. \eqref{two_comp_eta} reduces to:
\begin{equation}
    \eta=\frac{T}{10}\frac{\gamma_1^2c_{22}+\gamma_2^2c_{11}-2\gamma_1\gamma_2c_{12}}{c_{11}c_{22}-c_{12}^2}\to\frac{T}{10}\frac{(c_1\gamma_{0,1})^2(c_2^2c_{00,2})+(c_2\gamma_{0,2})^2(c_1^2c_{00,1})-0}{(c_1^2c_{00,1})(c_2^2c_{00,2})-0}=\frac{T}{10}\left(\frac{\gamma_{0,1}^2}{c_{00,1}}+\frac{\gamma_{0,2}^2}{c_{00,2}}\right)=\eta_1+\eta_2
    \label{eta_reduction_sigma0}
\end{equation}
This may lead to think that, if we consider $\sigma_{12}= 0$, that is, a system of two fluids which do not interact one with the other but only within themselves, then the viscosity of the binary mixture is the sum of the two. However, the issue is quite trickier than that, for both mathematical and physical reasons.\\

Indeed, let us note that the result \eqref{eta_reduction_sigma0}, in the limit $\sigma_{12}\to 0$, does not depend on the relative concentrations of the two components. This means that, if we now perform also the limit $x_1\to 0$ (without loss of generality) on \eqref{eta_reduction_sigma0}, we simply get:
\begin{equation}
    \lim_{x_1\to 0} \lim_{\sigma_{12}\to 0}\eta=\eta_1+\eta_2.
\label{eta_reduction_sigma0first_c10second}
\end{equation}
This implies that, even for infinitesimally small concentration of one of the two components of the mixture, the viscosity will differ by a finite amount from the contribution of the remaining component only. This turns out already counterintuitive and nonphysical.

A further insight into the issue is given if one reverses the order of the limit procedures in \eqref{eta_reduction_sigma0first_c10second}. In particular, let us first see how each term in \eqref{two_comp_eta} behaves when $x_1\to0$:\footnote{Here ``$a\sim o(x_1^b)$'' means ``$a$ goes to zero as $x_1^b$".}
\begin{align}
\gamma_1^2c_{22}\sim o(x_1^2),~~~\gamma_2^2c_{11}\sim x_2^2 \gamma_{0,2}^2\tilde{c}_{11}\sim o(x_1),~~~2\gamma_1\gamma_2c_{12}\sim o(x_1^2),~~~c_{11}c_{22}\sim \tilde{c}_{11}x_2^2 c_{00,2}\sim o(x_1),~~~c_{12}^2\sim o(x_1^2).
    \label{limits_x1to0}
\end{align}
This means that if we consider \eqref{two_comp_eta} when $x_1\to 0$, only the terms $o(x_1)$ are relevant and we get:
\begin{equation}
\eta\to \frac{T}{10}\frac{x_2^2 \gamma_{0,2}^2\tilde{c}_{11}}{\tilde{c}_{11}x_2^2 c_{00,2}}=\frac{T}{10}\frac{\gamma_{0,2}^2}{c_{00,2}}=\eta_2.
     \label{eta_reduction_c10}
\end{equation}
This is of course expected, since when $x_1\to 0$ only the second component is left in the mixture. However, this implies that now there is no dependence on $\sigma_{12}$, and if we perform the limit $\sigma_{12}\to 0$ we get:
\begin{equation}
\lim_{\sigma_{12}\to 0} \lim_{x_1\to 0}\eta=\eta_2.
\label{eta_reduction_c10first_sigma0second}
\end{equation}
What we observe is that Eqs. \eqref{eta_reduction_sigma0first_c10second} and \eqref{eta_reduction_c10first_sigma0second} lead to different results, and it is likely that performing the limits $\sigma_{12},x_1\to 0$ along any line $\sigma_{12}/x_1=$ constant will lead to values still different from the above.\\

The issue here is likely to come from the Chapman-Enskog approach itself. Indeed, a system with $\sigma_{12}=0$ is not in an equilibrated state, since the two components of a mixture cannot interact with one another, and the relaxation time of the system is $\tau_{\text{relax}}=+\infty$. This means that the Chapman-Enskog approximation is not applicable when $\sigma_{12}=0$, since there is no equilibrium state to expand from. As soon as $\sigma_{12}$ is however small, but not zero (as in the studies we performed in Section \ref{Results_averages}), the relaxation time is large but finite, and the Chapman-Enskog approximation is valid.\\

A similar problem occurs in the opposite limit $\sigma_{12}\to +\infty$. In particular, one can easily see that:
\begin{align}
\lim_{x_1\to 0} \lim_{\sigma_{12}\to +\infty}\eta&=0,
\label{eq:limits_sigmainfity_AppendixC_1}\\
\lim_{\sigma_{12}\to +\infty} \lim_{x_1\to 0}\eta&=\eta_2.
\label{eq:limits_sigmainfity_AppendixC_2}
\end{align}
It is likely that, as before, in this limit the CE approximation cannot be employed and therefore the formula \eqref{two_comp_eta} for $\eta$ breaks down.

\section{Scattering matrices in the Quasi Particle Model}
\label{Appendix D}

In the configuration which we label in the main text as \textit{QPM, anisotropic}, we are interested in studying the QCD bulk partons interacting with one another. In order to achieve this we calculate the gluon-gluon, gluon-quark, and quark-quark scattering matrices $\mathcal{M}$ using perturbative QCD (pQCD) Feynman rules at tree level.\\

In the QPM framework, both quarks and gluons are massive: this means that, in order to calculate the matrix element $\mathcal{M}$ corresponding to a scattering in QPM, we have to modify the expressions for the propagators as in \cite{Moreau:2019vhw}:
\begin{fmffile}{propagators}
\begin{align}
    \parbox{29mm}{\begin{fmfgraph*}(60,40)
    \fmfleft{i1}
    \fmfright{i2}
    \fmf{gluon, width=0.5, label=$q$, label.dist=10pt, label.side=top}{i1,i2}
    \fmflabel{$\mu, a$}{i1}
    \fmflabel{$\nu, b$}{i2}
\end{fmfgraph*}}&=-\delta_{ab}\frac{g^{\mu\nu}-q^\mu q^\nu/M_g^2}{q^2-M_g^2},\label{1.6}\\
\parbox{25mm}{\begin{fmfgraph*}(60,40)
    \fmfleft{i1}
    \fmfright{i2}
    \fmf{fermion, width=0.5, label=$q$, label.side=top}{i1,i2}
    \fmflabel{$i$}{i1}
    \fmflabel{$j$}{i2}
\end{fmfgraph*}}&=-\delta_{ij}\frac{\slashed{q}+M_q}{q^2-M_q^2}.
\label{1.7}
\end{align}
\end{fmffile}

That is, the usual pQCD propagator for the gluon has to be replaced with a massive vector propagator. In the above expressions $q$ is the 4-momentum of the exchanged particle, whereas the delta functions (over $a,b$ for the gluon, over $i,j$ for the quark in the above diagrams) ensure that the exchanged gluon/quark is connected with the other parts of the diagram carrying the same colour.

The addition of a mass in the gluon propagator, as in \eqref{1.6}, is not the only modification we have to make with respect to a standard pQCD Feynman diagram study. Indeed, while evaluating $|\mathcal{M}|^2$, also the sum over polarizations $\lambda$ of the gluons has to be modified as:
\begin{equation}
    \sum_\lambda \epsilon_\mu^\lambda(k)\epsilon_\nu^\lambda(k)=-g_{\mu\nu}+\frac{k_\mu k_\nu}{k^2},
    \label{eq:sum_over_polarizations_QPM}
\end{equation}
whereas in usual pQCD only the $-g_{\mu\nu}$ term in the right hand side would be present.\\

Once we discussed these technical subtleties, in this Appendix we now give more details on the calculation of the matrix elements used to evaluate the cross sections for massive partons. In order to fix the notation, let us remind that, by calling $(k_i,p_i)$ the initial 4-momenta of the particles and $(k_f,p_f)$ their final 4-momenta, the three Mandelstam variables are given by:
\begin{equation}
    s=(k_i+p_i)^2=(k_f+p_f)^2,~~~~~t=(k_i-k_f)^2=(p_i-p_f)^2,~~~~~u=(k_i-p_f)^2=(p_i-k_f)^2.
    \label{eq:mandelstam_variables}
\end{equation}
Moreover, the generators of SU(3) associated with QCD are the Gell-Mann matrices divided by two, i.e. $T^a=\lambda^a/2$, being $a=1,\dots, 8$ the gluon colour index \cite{Gell-Mann:1962yej}. The Lie algebra which the generators $T^a$ have to obey is given by the following commutation relations:
\begin{equation}
    [T^a,T^b]=if^{abc}T^c,
    \label{eq:lie_algebra_su3}
\end{equation}
where $f^{abc}$ are the SU(3) structure constants. The rules for the sum over colors which we are going to perform are given in detail in \cite{MacFarlane:1968vc}. Furthermore, the Dirac gamma matrices will be denoted by $\gamma^\mu$. The Dirac spinors for particles and anti-particles will be indicated by $u$ and $v$, respectively, and $\epsilon$ will denote the polarization vector for gluons.\\

We are going to proceed by listing all the matrix elements $\mathcal{M}$ of our interest, as derived from the Feynman rules applied to the proper diagrams. The final analytical expressions for the squared invariant matrix element $|\mathcal{M}|^2$ are then determined by averaging over initial spins and colors and summing over final ones. This has been done using the \textsc{feyncalc} package within \texttt{Wolfram Mathematica} \cite{Mertig:1990an,Shtabovenko:2016sxi,Shtabovenko:2020gxv}.

\subsection*{Quark-quark scattering}
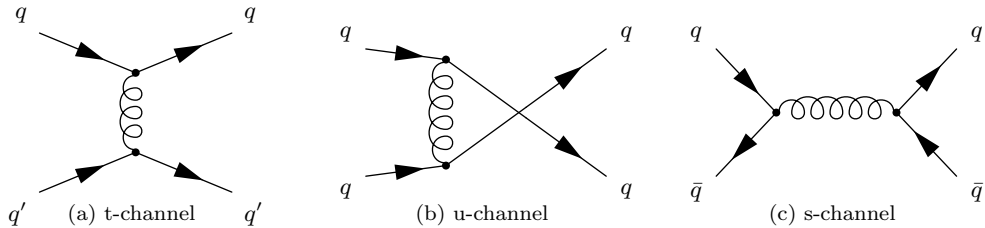
\begin{figure*}[ht]
\centering
\subfloat[width=.30\textwidth][{t-channel}]{
\begin{fmfgraph*}(90,60)
    \fmfleft{i1,i2}
    \fmfright{o1,o2}
    \fmf{gluon, width=0.5}{v1,v2}
    \fmf{fermion, width=0.5}{i2,v1}
    \fmf{fermion, width=0.5}{v1,o2}
    \fmf{fermion, width=0.5}{i1,v2}
    \fmf{fermion,  width=0.5}{v2,o1}
    \fmfset{dot_size}{0.7mm} 
    \fmfdot{v1,v2}
    \fmflabel{$q'$}{i1}
    \fmflabel{$q$}{i2}
    \fmflabel{$q'$}{o1}
    \fmflabel{$q$}{o2}
\end{fmfgraph*}} \hspace{35pt}
\subfloat[width=.30\textwidth][{u-channel}]{
\begin{fmfgraph*}(90,60)
    \fmftop{i1,i2}
    \fmfbottom{o1,o2}
    \fmf{fermion, tension=2, width=0.5}{i1,v2}
    \fmf{gluon, tension=0.3, width=0.5}{v2,w2}
    \fmf{fermion, tension=2, width=0.5}{o1,w2}
    \fmf{phantom}{w2,o2}
    \fmf{phantom}{i2,v2}
    \fmffreeze
    \fmf{plain, width=0.5}{v2,v22}
     \fmf{plain, width=0.5}{w2,w22}
    \fmf{fermion, width=0.5}{v22,o2}
    \fmf{fermion, width=0.5}{w22,i2}
    \fmfset{dot_size}{0.7mm} 
    \fmfdot{w2,v2}
    \fmflabel{$q$}{i1}
    \fmflabel{$q$}{i2}
    \fmflabel{$q$}{o1}
    \fmflabel{$q$}{o2}
\end{fmfgraph*}} 
\hspace{35pt}
\subfloat[width=.30\textwidth][{s-channel}]{
\begin{fmfgraph*}(90,60)
    \fmftop{i1,i2}
    \fmfbottom{o1,o2}
    \fmf{gluon, width=0.5}{v1,v2}
    \fmf{fermion, width=0.5}{o2,v1,i2}
    \fmf{fermion, width=0.5}{i1,v2,o1}
    \fmfset{dot_size}{0.7mm} 
    \fmfdot{v1,v2}
    \fmflabel{$q$}{i1}
    \fmflabel{$q$}{i2}
    \fmflabel{$\Bar{q}$}{o1}
    \fmflabel{$\Bar{q}$}{o2}
\end{fmfgraph*}}
\caption{Leading-order Feynman diagrams of the quark-quark scatterings.}
\label{qq_FD}
\end{figure*}

When dealing with quark-quark scattering, depending on their flavours we may have the presence of not only the direct ($t$-channel) diagram, but also of the exchange or of the annihilation diagram ($u$- and $s$-channels, respectively). The diagrams involved in the quark-quark scattering are depicted in Figure \ref{qq_FD}.
\begin{fmffile}{quarkquark}
\vspace{1em}
\end{fmffile}

By denoting as $\alpha,\beta,\gamma,\delta$ the flavour indices, $i,j,k,l=1,2,3$ the quark colour indices and $a,b,c,d,e=1,\dots,8$ the gluon colour indices, each of the diagrams in Figure \ref{qq_FD} carries the following contribution for the $\mathcal{M}$ matrix \cite{Moreau:2019vhw}:
\begin{align}
i\mathcal{M}_t(q_a^iq_\beta^k\to q_\delta^j q_\gamma^l)&=\delta_{\alpha\delta}\delta_{\beta\gamma}\Bar{u}_\delta^j (k_f)(-ig\gamma^\mu T_{ij}^a)u_\alpha^i(k_i)\left[-i\frac{g_{\mu\nu}-(q_\mu^tq_\nu^t)/M_g^2}{(k_f-k_i)^2-M_g^2}\right]\Bar{u}_\gamma^l(p_f)(-ig\gamma^\nu T_{kl}^a)u_\beta^k(p_i),\nonumber\\
i\mathcal{M}_u(q_a^iq_\beta^k\to q_\delta^j q_\gamma^l)&=-\delta_{\alpha\delta}\delta_{\alpha\delta}\delta_{\beta\gamma}\Bar{u}_\delta^j (k_f)(-ig\gamma^\nu T_{kj}^a)u_\beta^k(p_i)\left[-i\frac{g_{\mu\nu}-(q_\mu^uq_\nu^u)/M_g^2}{(p_f-k_i)^2-M_g^2}\right]\Bar{u}_\gamma^l(p_f)(-ig\gamma^\mu T_{il}^a)u_\alpha^i(k_i),\nonumber\\
i\mathcal{M}_s(q_a^iq_\beta^k\to q_\delta^j q_\gamma^l)&=-\delta_{\alpha\Bar{\beta}}\delta_{\delta \Bar{\gamma}}\Bar{u}_\delta^j (k_f)(-ig\gamma^\nu T_{lj}^a)v_\gamma^l(p_f)\left[-i\frac{g_{\mu\nu}-(q_\mu^sq_\nu^s)/M_g^2}{(k_i+p_i)^2-M_g^2}\right]\Bar{v}_\beta^k(p_i)(-ig\gamma^\mu T_{ik}^a)u_\alpha^i(k_i),
\label{eq:qq_scattering_matrices_appendixD}
\end{align}
where $q^\mu_t=(k_f-k_i)^\mu$, $q^\mu_u=(p_i-k_f)^\mu$ and $q_s^\mu=(k_i+p_i)^\mu$ is the momentum of the exchanged gluon in each case. These matrices are then added up, modulus-squared and averaged (summed) over initial (final) colors and spins, that is:
\begin{equation}
    |\mathcal{M}(q_\alpha q_\beta\to q_\delta q_\gamma)|^2=\frac{1}{3\times 2}\frac{1}{3\times 2}\sum_{\text{colour}}\sum_{\text{spin}}|\mathcal{M}_t+\mathcal{M}_u+\mathcal{M}_s|^2.
    \label{eq:squared_modulus_AppendixD}
\end{equation}
The action of the Kronecker deltas in the expressions \eqref{eq:qq_scattering_matrices_appendixD} for $\mathcal{M}$ therefore selects the diagrams of our interest depending on the quarks involved. In practice, for quarks of different flavours we are going to consider only the $t$-channel diagram, for $q\Bar{q}\to q\Bar{q}$ we are considering both the $t$- and the $s$-channel diagrams and for $qq\to qq$ we are considering both the $t$- and the $u$-channel diagrams.

The final results are:

\begin{align}
    |\mathcal{M}(qq'\to qq')|^2&=\frac{4g^4}{9(M_g^2-t)^2}\left[2M_q^4+2M_q^2(2M_{q'}^2-s+t-u)+2M_{q'}^4-2M_{q'}^2(s-t+u)+s^2+u^2\right],\label{eq_Mqq'_AppendixD}\\
|\mathcal{M}(qq\to qq)|^2&=\frac{4g^4}{27(M_g^2-t)^2(M_g^2-u)^2}\{8M_q^4[5M_g^4-5M_g^2(t+u)+3(t^2+u^2)-tu]-4M_q^2[M_g^4(3s+t+u)-\nonumber\\
&M_g^2(3s(t+u)+7t^2-10tu+7u^2)+3s(t^2+u^2)-3stu+(t+u)(3(t-u)^2+tu)]+\nonumber\\
&M_g^4(4s^2+3(t^2+u^2))-\left.2M_g^2(t+u)(2s^2+3(t^2-tu+u^2))+3s^2t^2-2s^2tu+3s^2u^2+3t^4+3u^4\right\},
\label{eq_Mqq_AppendixD}
\end{align}
\begin{align}
|\mathcal{M}(q\Bar{q}\to q\Bar{q})|^2&=\frac{4g^4}{27(M_g^2-s)^2(M_g^2-t)^2}\left\{8M_q^4[5M_g^4-5M_g^2(s+t)+3(s^2+t^2)-st]-4M_q^2[M_g^4(3u+s+t)-\right.\nonumber\\
&M_g^2(3u(s+t)+7s^2-10st+7t^2)+3u(s^2+t^2)-3stu+(s+t)(3(s-t)^2+st)]+\nonumber\\
&M_g^4(4u^2+3(s^2+t^2))-\left.2M_g^2(s+t)(2u^2+3(s^2-st+t^2))+3u^2s^2-2u^2ts+3u^2t^2+3s^4+3t^4\right\}.
\label{eq_Mqqbar_AppendixD}
\end{align}
\subsection*{Quark-gluon scattering}
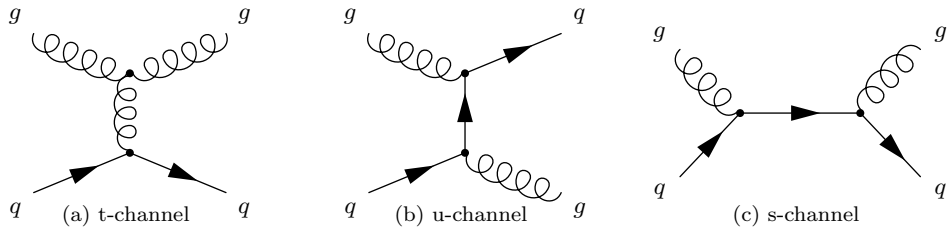
\begin{figure*}[ht]
\begin{minipage}[ht]{0.24\textwidth}
\centering
\subfloat[width=.30\textwidth][{t-channel}]{
\begin{fmfgraph*}(90,60)
    \fmfleft{i1,i2}
    \fmfright{o1,o2}
    \fmf{gluon, width=0.5}{v1,v2}
    \fmf{gluon, width=0.5}{i2,v1}
    \fmf{gluon, width=0.5}{v1,o2}
    \fmf{fermion, width=0.5}{i1,v2}
    \fmf{fermion, width=0.5}{v2,o1}
    \fmfset{dot_size}{0.7mm} 
    \fmfdot{v1,v2}
    \fmflabel{$q$}{i1}
    \fmflabel{$g$}{i2}
    \fmflabel{$q$}{o1}
    \fmflabel{$g$}{o2}
\end{fmfgraph*}}
\end{minipage}
\begin{minipage}[ht]{0.24\textwidth}
\centering
\subfloat[width=.30\textwidth][{u-channel}]{
\begin{fmfgraph*}(90,60)
    \fmfleft{i1,i2}
    \fmfright{o1,o2}
    \fmf{fermion, width=0.5}{v2,v1}
    \fmf{gluon, width=0.5}{i2,v1}
    \fmf{fermion, width=0.5}{v1,o2}
    \fmf{fermion, width=0.5}{i1,v2}
    \fmf{gluon, width=0.5}{v2,o1}
    \fmfset{dot_size}{0.7mm} 
    \fmfdot{v1,v2}
    \fmflabel{$q$}{i1}
    \fmflabel{$g$}{i2}
    \fmflabel{$g$}{o1}
    \fmflabel{$q$}{o2}
\end{fmfgraph*}}
\end{minipage}
\begin{minipage}[ht]{0.24\textwidth}
\centering
\subfloat[width=.30\textwidth][{s-channel}]{
\begin{fmfgraph*}(90,60)
    \fmftop{i1,i2}
    \fmfbottom{o1,o2}
    \fmf{fermion, width=0.5}{o1,v1,v2,o2}
    \fmf{gluon, width=0.5}{i1,v1}
    \fmf{gluon, width=0.5}{i2,v2}
    \fmfset{dot_size}{0.7mm} 
    \fmfdot{v1,v2}
    \fmflabel{$g$}{i1}
    \fmflabel{$g$}{i2}
    \fmflabel{$q$}{o1}
    \fmflabel{$q$}{o2}
\end{fmfgraph*}}
\end{minipage}
\caption{Leading-order Feynman diagrams of the quark-gluon scatterings.}
\label{gq_FD}
\end{figure*}

The diagrams involved in the $qg\to qg$ scattering are depicted in Figure \ref{gq_FD}.

\begin{fmffile}{quarkgluon}
\vspace{1em}
\end{fmffile}
These diagrams correspond to the following expressions for the invariant matrix elements \cite{Moreau:2019vhw}:
\begin{align}
i\mathcal{M}_t(g^a q^i\to g^b q^j)&=(\epsilon_{b,f}^*)_\nu[-g f^{cab}C^{\lambda \mu\nu}(k_i-k_f,-k_i,k_f)](\epsilon_{a,i})_\mu\left[-i\frac{g_{\lambda \tau}-(q_\lambda^tq_\tau^t)/M_g^2}{(k_f-k_i)^2-M_g^2}\right]\Bar{u}^j(p_f)(-ig\gamma^\tau T_{ij}^c)u^i(p_i),\nonumber\\
i\mathcal{M}_u(g^a q^i\to g^b q^j)&=\Bar{u}^j(p_f)(-ig\gamma^\mu T_{kj}^a))(\epsilon_{a,i})_\mu\left[i\frac{\slashed{q}^u+M_q}{u-M_q^2}\right](\epsilon_{b,f}^*)_\nu (-ig\gamma^\nu T_{ik}^b)u^i(p_i),\nonumber\\
i\mathcal{M}_s(g^a q^i\to g^b q^j)&=\Bar{u}^j(p_f)(-ig\gamma^\nu T_{lj}^b))(\epsilon_{b,f}^*)_\nu\left[i\frac{\slashed{q}^s+M_q}{s-M_q^2}\right](\epsilon_{a,i})_\mu (-ig\gamma^\mu T_{il}^a)u^i(p_i),
\label{eq:qg_scattering_matrices_appendixD}
\end{align}
with the 3-gluon vertex being (the 3-momenta are considered as all entering the vertex):
\begin{equation}
    C^{\lambda\mu\nu}(q_1,q_2,q_3)=[(q_1-q_2)^\nu g^{\lambda\mu}+(q_2-q_3)^\lambda g^{\mu\nu}+(q_3-q_1)^\mu g^{\lambda \nu}].
    \label{eq:3_gluon_vertex_appendixD}
\end{equation}
The quantity of our interest is obtained by averaging over the initial - and summing over the final - spin and colors, that is
\begin{equation}
    |\mathcal{M}(gq\to gq)|^2=\frac{1}{8\times 2}\frac{1}{3\times 2}\sum_{\text{colour}}\sum_{\text{spin}}|\mathcal{M}_t+\mathcal{M}_u
+\mathcal{M}_s|^2.
\label{eq_Mqg_formula_AppendixD}
\end{equation}
The final result is
\begin{align}
    |\mathcal{M}(&qg\to qg)|^2=\nonumber\\
    g^4&\{207M_q^{12}+9M_q^{10}[8M_g^2-21(s+u)]-\nonumber\\
M_q^8&[255M_g^4-8M_g^2(45s+3t+45u)+27s^2+585su+12t^2+27u^2]+\nonumber\\
9M_q^6&[-8M_g^6+82M_g^4(s+u)-4M_g^2(11s^2+62su+11u^2)+17s^3+105s^2u+105su^2+17u^3]+\nonumber\\
M_q^4&[48M_g^8+4M_g^6(43s-6t+43u)+M_g^4(-291s^2+4s(4t-551u)+12t^2+16tu-291u^2)+\nonumber\\
&4M_g^2(54s^3+s^2(450u-3t)-2s(t^2+7tu-225u^2)+u(-2t^2-3tu+54u^2))-\nonumber\\
&90s^4-603s^3u+6s^2t^2-693s^2u^2+28st^2u-603su^3+6t^2u^2-90u^4]+\nonumber\\
M_q^2&[-48M_g^8(s+u)+4M_g^6(9s^2+2s(3t-68u)+u(6t+9u))+2M_g^4(8s^3+533s^2u-s(6t^2+32tu-533u^2)-\nonumber\\
&6t^2u-8u^3)+4M_g^2(9s^4-s^3(t-189u)+s^2u(162u-7t)+su(-8t^2-7tu+189u^2)+u^3(9u-t))+\nonumber\\
&(s+u)(18s^4+144s^3u+s^2(63u^2-2t^2)-4s(3t^2u-36u^3)-2t^2u^2+18u^4)]+\nonumber\\
2M_g^8&(8s^2+8su+8u^2)-4M_g^6(9s^3-s^2(t+34u)+2su(4t-17u)+u^2(9u-t))+M_g^4(18s^4-232s^3u+\nonumber\\
&s^2(-2t^2+16tu-171u^2)+2su(8t^2+8tu-116u^2)-2t^2u^2+18u^4)+4M_g^2su(27s^3-s^2(t-27u)+s(27u^2-2t^2)+\nonumber\\
&u(-2t^2-tu+27u^2))+su(-18s^4+s^2(2t^2-45u^2)+2u^2(t^2-9u^2))\}/[36(M_q^2-s)^2(M_q^2-u)^2(M_g^2-t)^2].
\label{eq_Mqg_result_AppendixD}
\end{align}

\subsection*{Gluon-gluon scattering}
\begin{figure*}[ht]
\begin{fmffile}{gluongluon}
\vspace{1em}
\begin{minipage}[ht]{0.24\textwidth}
\centering
\subfloat[width=.30\textwidth][{t-channel}]{
\begin{fmfgraph*}(90,60)
    \fmfleft{i1,i2}
    \fmfright{o1,o2}
    \fmf{gluon, width=0.5}{v1,v2}
    \fmf{gluon, width=0.5}{i2,v1}
    \fmf{gluon, width=0.5}{v1,o2}
    \fmf{gluon, width=0.5}{i1,v2}
    \fmf{gluon, width=0.5}{v2,o1}
    \fmfset{dot_size}{0.7mm} 
    \fmfdot{v1,v2}
    \fmflabel{$g$}{i1}
    \fmflabel{$g$}{i2}
    \fmflabel{$g$}{o1}
    \fmflabel{$g$}{o2}
\end{fmfgraph*}}
\end{minipage}
\begin{minipage}[ht]{0.24\textwidth}
\centering
\subfloat[width=.30\textwidth][{u-channel}]{
\begin{fmfgraph*}(90,60)
    \fmftop{i1,i2}
    \fmfbottom{o1,o2}
    \fmf{gluon, tension=2, width=0.5}{i1,v2}
    \fmf{gluon, tension=0.3, width=0.5}{v2,w2}
    \fmf{gluon, tension=2, width=0.5}{o1,w2}
    \fmf{phantom}{w2,o2}
    \fmf{phantom}{i2,v2}
    \fmffreeze
    \fmf{gluon, width=0.5}{v2,o2}
    \fmf{gluon, width=0.5}{w2,i2}
    \fmfset{dot_size}{0.7mm} 
    \fmfdot{w2,v2}
    \fmflabel{$g$}{i1}
    \fmflabel{$g$}{i2}
    \fmflabel{$g$}{o1}
    \fmflabel{$g$}{o2}
\end{fmfgraph*}}
\end{minipage}
\begin{minipage}[ht]{0.24\textwidth}
\centering
\subfloat[width=.30\textwidth][{s-channel}]{
\begin{fmfgraph*}(90,60)
    \fmftop{i1,i2}
    \fmfbottom{o1,o2}
    \fmf{gluon, width=0.5}{v1,v2}
    \fmf{gluon, width=0.5}{i2,v1}
    \fmf{gluon, width=0.5}{v1,o2}
    \fmf{gluon, width=0.5}{i1,v2}
    \fmf{gluon, width=0.5}{v2,o1}
    \fmfset{dot_size}{0.7mm} 
    \fmfdot{v1,v2}
    \fmflabel{$g$}{i1}
    \fmflabel{$g$}{i2}
    \fmflabel{$g$}{o1}
    \fmflabel{$g$}{o2}
\end{fmfgraph*}}
\end{minipage}
\begin{minipage}[ht]{0.24\textwidth}
\centering
\subfloat[width=.30\textwidth][{4-point}]{
\begin{fmfgraph*}(90,60)
    \fmfleft{i1,i2}
    \fmfright{o1,o2}
    \fmf{gluon, width=0.5}{i2,v1}
    \fmf{gluon, width=0.5}{v1,o2}
    \fmf{gluon, width=0.5}{i1,v1}
    \fmf{gluon, width=0.5}{v1,o1}
    \fmfset{dot_size}{0.7mm} 
    \fmfdot{v1}
    \fmflabel{$g$}{i1}
    \fmflabel{$g$}{i2}
    \fmflabel{$g$}{o1}
    \fmflabel{$g$}{o2}
\end{fmfgraph*}}
\end{minipage}
\caption{Leading-order Feynman diagrams of the gluon-gluon scatterings.}
\label{gg_FD}
\end{fmffile}
\end{figure*}
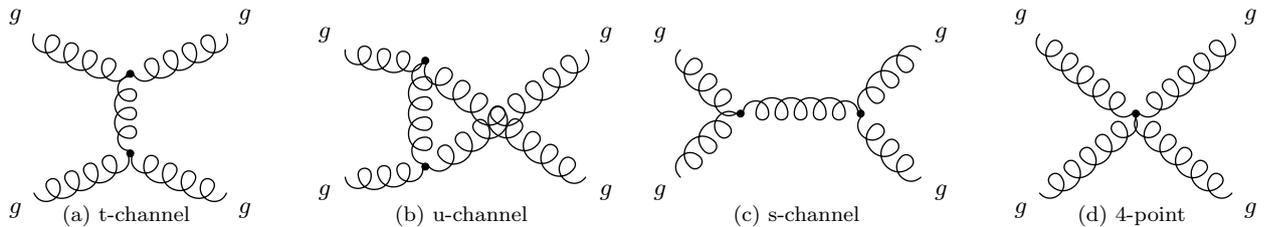

The diagrams involved in the $gg\to gg$ scattering are depicted in Figure \ref{gg_FD}.

Such diagrams lead to the following expressions for the invariant matrix elements \cite{Moreau:2019vhw}:
\begin{align}
i\mathcal{M}_t(g^a g^b \to g^c g^d)=(\epsilon_{d,4}^*)_\sigma[-g f^{ead}&C^{\tau\lambda\sigma}(q_1-q_4,-q_1,q_4)](\epsilon_{a,1})_\lambda\cdot \nonumber\\
&\left[-i\frac{g_{\tau \tau'}-(q_\tau^tq_{\tau'}^t)/M_g^2}{(q_4-q_1)^2-M_g^2}\right](\epsilon_{c,3}^*)_\nu [-g f^{ecb}C^{\tau'\nu\mu}(-q_3+q_2,q_3,-q_2)](\epsilon_{b,2})_\mu,\nonumber\\
i\mathcal{M}_s(g^a g^b \to g^c g^d)=(\epsilon_{d,4}^*)_\sigma[-g f^{edc}&C^{\tau'\sigma \nu}(q_4-q_3,q_4,q_3)](\epsilon_{c,3}^*)_\nu\cdot\nonumber\\
&\left[-i\frac{g_{\tau \tau'}-(q_\tau^sq_{\tau'}^s)/M_g^2}{(q_1+q_2)^2-M_g^2}\right](\epsilon_{b,2})_\mu [-g f^{eba}C^{\tau\mu\lambda}(q_2+q_1,-q_2,-q_1)](\epsilon_{a,1})_\lambda,\nonumber\\
i\mathcal{M}_u(g^a g^b \to g^c g^d)=(\epsilon_{d,4}^*)_\sigma[-g f^{edb}&C^{\tau'\sigma \mu}(-q_4+q_2,q_4,-q_2)](\epsilon_{b,2})_\mu\cdot\nonumber\\
&\left[-i\frac{g_{\tau \tau'}-(q_\tau^uq_{\tau'}^u)/M_g^2}{(q_2-q_4)^2-M_g^2}\right](\epsilon_{c,3}^*)_\nu [-g f^{eac}C^{\tau\lambda\nu}(q_1-q_3,-q_1,q_3)](\epsilon_{a,1})_\lambda,\nonumber\\
i\mathcal{M}_4(g^a g^b \to g^c g^d)=-ig^2[f^{abe}f^{cde}(&g^{\lambda \nu}g^{\mu \sigma}-g^{\lambda \sigma}g^{\mu\nu})+\nonumber\\
f^{ace}f^{bde}(g^{\lambda \mu}&g^{\nu \sigma}-g^{\lambda \sigma}g^{\mu\nu})+f^{ade}f^{bce}(g^{\lambda \mu}g^{\sigma \nu}-g^{\lambda \nu}g^{\sigma\mu})](\epsilon_{d,4}^*)_\sigma(\epsilon_{c,3}^*)_\nu(\epsilon_{b,2})_\mu(\epsilon_{a,1})_\lambda.
\label{eq:gg_scattering_matrices_appendixD}
\end{align}

The invariant matrix element, after proper averaging over the initial - and summing over the final - gluon states is
\begin{equation}
    |\mathcal{M}(gg\to gg)|^2=\frac{1}{8\times 2}\frac{1}{8\times 2}\sum_{\text{colour}}\sum_{\text{pol.}}|\mathcal{M}_t+\mathcal{M}_u
+\mathcal{M}_s+\mathcal{M}_4|^2.
\label{eq_Mgg_formula_AppendixD}
\end{equation}

The final result is:
\begin{align}
|\mathcal{M}(&gg\to gg)|^2=\nonumber \\
   9g^4&[56160M_g^{16}-100224M_g^{14}(t+u)+
   48M_g^{12}(2111t^2+1934tu+2111u^2)-
1208M_g^{10}(49t^3+53t^2u+53tu^2+49u^3)+\nonumber\\
M_g^8&(20559t^4+30182t^3u+34705t^2u^2+30182tu^3+20559u^4)-\nonumber\\
M_g^6&(3948t^5+8743t^4u+12266t^3u^2+12266t^2u^3+8743tu^4+3948u^5)+\nonumber\\
M_g^4&(329t^6+1059t^5u+2035t^4u^2+2289t^3u^3+2035t^2u^4+1059tu^5+329u^6)-\nonumber\\
2M_g^2&tu(3t^5+15t^4u+29t^3u^2+29t^2u^3+15tu^4+3u^5)+\nonumber\\
&~~~~~t^2u^2(t+u)^2(t^2+tu+u^2)]/[512M_g^4(M_g^2-s)^2(M_g^2-t)^2(M_g^2-u)^2].
\label{eq_Mgg_result_AppendixD}
\end{align}

\section{Tables on masses and cross sections}
\label{Appendix E}
In this Appendix we list the specific values and relations used to reproduce the results shown in the Figures. In particular, for each case we list the masses used for the two species, the coupling constant, and the squared scattering matrices $|\mathcal{M}|^2$ (those are related to the differential cross section by
$d \sigma/d\Omega=|\mathcal{M}|^2/64\pi^2 s$).\\

In the \textit{massless, isotropic} case (fourth entry of Table \ref{table:paramters_fig1}), the quantities we are referring to are textbook massless pQCD scattering matrices \cite{Ellis:1996mzs}:
\begin{align}
    |\mathcal{M}_{gg}|^2&=g^4\left[\frac{9}{2}\left(3-\frac{ut}{s^2}-\frac{su}{t^2}-\frac{st}{u^2}\right)\right].\label{eq:Mgg_AppendixE}\\
    |\mathcal{M}_{gq}|^2&=g^4\left[\frac{s^2+u^2}{t^2}-\frac{4}{9}\frac{s^2+u^2}{su}\right],\label{eq:Mgq_AppendixE}\\
    |\mathcal{M}_{qq}|^2&=g^4\left[\frac{4}{9}\left(\frac{s^2+u^2}{t^2}+\frac{s^2+t^2}{u^2}\right)-\frac{8}{27}\frac{s^2}{ut}\right],\label{eq:Mqq_AppendixE}\\
    |\mathcal{M}_{q\bar{q}}|^2&=g^4\left[\frac{4}{9}\left(\frac{s^2+u^2}{t^2}+\frac{t^2+u^2}{s^2}\right)-\frac{8}{27}\frac{u^2}{st}\right],\label{eq:Mqqbar_AppendixE}\\
    |\mathcal{M}_{qq'}|^2&=g^4\left[\frac{4}{9}\left(\frac{s^2+u^2}{t^2}\right)\right],\label{eq:Mqq'_AppendixE}
\end{align}

The factors appearing in the expression for $|\mathcal{M}_{22}|^2$ in the cases \textit{massless, anisotropic} and \textit{QPM, anisotropic} refer to the probability that in a collision among two quarks we have a pair of equal quarks, a quark-antiquark pair and all the other possible cases, respectively. Those have been derived via basic combinatorics, supposing a very large total number of particles $N$, as:\footnote{Since the relative abundance in the mixture only depends on the mass, all quarks and antiquarks appear in equal number. This implies that the above ‘probabilities' are fixed and do not depend on either temperature nor quark mass.}
\begin{align}
\mathrm{P}_{qq\to qq}=&\frac{\text{\# of possible $qq$ pairs}}{\text{total \# of possible pairs}}=\frac{N(N/6-1)/2}{N(N-1)/2}\to\frac16,\label{eq:Pqq}\\
    \mathrm{P}_{q\Bar{q}\to q\Bar{q}}=&\frac{\text{\# of possible $q\Bar{q}$ pairs}}{\text{total \# of possible pairs}}=\frac{N(N/6)/2}{N(N-1)/2}\to\frac16,\label{eq:Pqqbar}\\
    \mathrm{P}_{qq'\to qq'}=&1-\mathrm{P}_{qq\to qq}-\mathrm{P}_{q\Bar{q}\to q\Bar{q}}\to \frac{2}{3}.\label{eq:Pqq'}
\end{align}
This has to be done since our Chapman-Enskog formalism for the binary mixture is not able to distinguish among the different quark processes, since we can only deal with gluons and a unique flavour of quarks. Due to this fact, we have considered an ‘averaged’ quark-quark differential cross section. On the other hand, the transport code which has been employed to evaluate the
Green-Kubo correlator allows for the different quark cross sections to be implemented singularly for each different channel.\\

Furthermore, note that the values of $|\mathcal{M}|^2$ which have been chosen in the ‘isotropic' cases are just the prefactors appearing in Eq. \eqref{eq:Mgg_AppendixE} (i.e. $9/2$), Eq. \eqref{eq:Mgq_AppendixE} (i.e. $1-4/9=5/9$) and Eq. \eqref{eq:Mqq'_AppendixE} (i.e. $4/9$).\\

Finally, when dealing with the physical case of Fig. \ref{eta/s_averaged} (parameters in Table \ref{table:paramters_fig5}), note that the coupling constant $g(T)$ appearing in the QPM masses \eqref{3.9} is in principle not equal to the coupling arising from the Feynman diagram vertices. Indeed, to evaluate the shear viscosity in a physical case, we want to use the same coupling that gives heavy quark drag and diffusion coefficients in agreement with the experimental data for the nuclear modification factor $R_{AA}$ and the elliptic flows $v_n$ of open heavy flavor observables \cite{Scardina:2017ipo, Sambataro:2024mkr}. This correction corresponds to a multiplicative factor of 1.25 to the coupling constant.

\begin{table}[H]
\centering
\begin{tblr}{colspec={M{4cm}||M{1.8cm}|M{1cm}|M{1cm}|M{5.5cm}|M{1.5cm}|}, row{2-4}={8ex}}
  \textbf{Figure \ref{eta_over_T3_vs_temperature}} & $m_1, m_2$ &    $|\mathcal{M}_{11}|^2$   & $|\mathcal{M}_{12}|^2$    & $|\mathcal{M}_{22}|^2$  & coupling $g$   \\\hline\hline
 massless, isotropic   & 0,0 &  $g^4\,9/2$ &  $g^4\,5/9$ & $g^4\,4/9$ & 2\\\hline
constant masses, isotropic & 0.8, 0.5 GeV &  $g^4\,9/2$ &  $g^4\,5/9$ & $g^4\,4/9$ & 2\\\hline
 QPM, isotropic  & \makecell{$m_g, m_q$\\  \eqref{3.9}} &  $g^4\,9/2$ &  $g^4\,5/9$ & $g^4\,4/9$ & \makecell{$g(T)$\\ \eqref{eq:g_fit}}\\\hline
 massless, anisotropic  & 0, 0 &  \makecell{$|\mathcal{M}_{gg}|^2$\\  \eqref{eq:Mgg_AppendixE}} &  \makecell{$|\mathcal{M}_{qg}|^2$\\  \eqref{eq:Mgq_AppendixE}}& \makecell{$1/6|\mathcal{M}_{qq}|^2+1/6|\mathcal{M}_{q\bar{q}}|^2+2/3|\mathcal{M}_{qq'}|^2$ \\  \eqref{eq:Mqq_AppendixE}, \eqref{eq:Mqqbar_AppendixE}, \eqref{eq:Mqq'_AppendixE}} & 2\\\hline  
 QPM, anisotropic & \makecell{$m_g, m_q$\\  \eqref{3.9}}  &  \makecell{$|\mathcal{M}_{gg}|^2$ \\ \eqref{eq_Mgg_result_AppendixD}}&  \makecell{$|\mathcal{M}_{qg}|^2$\\  \eqref{eq_Mqg_result_AppendixD}}& 
\makecell{$1/6|\mathcal{M}_{qq}|^2+1/6|\mathcal{M}_{q\bar{q}}|^2+2/3|\mathcal{M}_{qq'}|^2$ \\  \eqref{eq_Mqq'_AppendixD}, \eqref{eq_Mqq_AppendixD}, \eqref{eq_Mqqbar_AppendixD}} & \makecell{$g(T)$\\ \eqref{eq:g_fit}}\\\hline 
\end{tblr}
\caption{Parameters used in Fig. \ref{eta_over_T3_vs_temperature}.}
 \label{table:paramters_fig1}
\end{table}

\begin{table}[H]
\centering
\begin{tblr}{colspec={M{5cm}||M{1.5cm}|M{1cm}|M{1cm}|M{1.5cm}|M{1.4cm}|}, row{2-6}={6ex}}
 \textbf{Figure \ref{eta_vs_M12squared}} & $m_1,m_2$ &    $|\mathcal{M}_{11}|^2$   & $|\mathcal{M}_{22}|^2$  & coupling $g$ & $T$  \\\hline\hline
massless, isotropic, $\sigma_{11}=\sigma_{22}$   & 0, 0 &  $g^4\,4/9$ &   $g^4\,4/9$ & 2& 0.5 GeV\\\hline
massless, isotropic, $\sigma_{11}\simeq 10\,\sigma_{22}$ & 0, 0 &  $g^4\,9/2$ &   $g^4\,4/9$ & 2& 0.5 GeV\\\hline
QPM, isotropic, $\sigma_{11}\simeq 10\,\sigma_{22}$  & \makecell{$m_g, m_q$ \\ \eqref{3.9}} &  $g^4\,9/2$ & $g^4\,4/9$ & 2 & 0.5 GeV\\\hline
\end{tblr}
\caption{Parameters used in Fig. \ref{eta_vs_M12squared}.}
 \label{table:paramters_fig2}
\end{table}

\begin{table}[H]
\centering
\begin{tblr}{colspec={M{5cm}||M{1.5cm}|M{1cm}|M{1cm}|M{1.5cm}|M{1.4cm}|}, row{2-6}={4ex}}
 \textbf{Figure \ref{eta_vs_concentrations}} & $m_1,m_2$ &    $|\mathcal{M}_{11}|^2$   & $|\mathcal{M}_{22}|^2$  & coupling $g$ & $T$  \\\hline\hline
massless, isotropic, $\sigma_{11}=\sigma_{22}$   & 0, 0 &  $g^4\,4/9$ &   $g^4\,4/9$ & 2& 0.5 GeV\\\hline
massless, isotropic, $\sigma_{11}\simeq 10\,\sigma_{22}$ & 0, 0 &  $g^4\,9/2$ &   $g^4\,4/9$ & 2& 0.5 GeV\\\hline
\end{tblr}
\caption{Parameters used in Fig. \ref{eta_vs_concentrations}.}
 \label{table:paramters_fig3}
\end{table}

\begin{table}[H]
\centering
\begin{tblr}{colspec={M{2.5cm}||M{1cm}|M{1.3cm}|M{1.3cm}|M{6.8cm}|M{1.5cm}|}, row{2-4}={9ex}}
 \textbf{Figure \ref{eta/s_averaged}} & $m_1, m_2$ &    $|\mathcal{M}_{11}|^2$   & $|\mathcal{M}_{12}|^2$    & $|\mathcal{M}_{22}|^2$  & coupling $g$   \\\hline\hline
\makecell{Full 2-Component\\ Chapman-Enskog} & \makecell{$m_g,m_q$ \vspace{0.2em}\\  \eqref{3.9}}  &  \makecell{$\frac{\displaystyle{|\mathcal{M}_{gg}|^2}}{\displaystyle{(1.25)^4}}$ \vspace{0.2em}\\ \eqref{eq_Mgg_result_AppendixD}}&  \makecell{$\frac{\displaystyle{|\mathcal{M}_{qg}|^2}}{\displaystyle{(1.25)^4}}$ \vspace{0.2em}\\\eqref{eq_Mqg_result_AppendixD}}& 
\makecell{$\left(1/6|\mathcal{M}_{qq}|^2+1/6|\mathcal{M}_{q\bar{q}}|^2+2/3|\mathcal{M}_{qq'}|^2\right)/(1.25)^4$ \vspace{0.2em}\\  \eqref{eq_Mqq'_AppendixD}, \eqref{eq_Mqq_AppendixD}, \eqref{eq_Mqqbar_AppendixD}} & \makecell{$g(T)$ \vspace{0.2em}\\\eqref{eq:g_fit}}\\\hline 
Green-Kubo & \makecell{$m_g,m_q$\vspace{0.2em}\\  \eqref{3.9}}  &  \makecell{$\frac{\displaystyle{|\mathcal{M}_{gg}|^2}}{\displaystyle{(1.25)^4}}$ \vspace{0.2em}\\ \eqref{eq_Mgg_result_AppendixD}}&  \makecell{$\frac{\displaystyle{|\mathcal{M}_{qg}|^2}}{\displaystyle{(1.25)^4}}$ \vspace{0.2em}\\\eqref{eq_Mqg_result_AppendixD}}& 
\makecell{$|\mathcal{M}_{qq'}|^2/(1.25)^4$ \eqref{eq_Mqq'_AppendixD} for $qq'\to qq'$\\
$|\mathcal{M}_{qq}|^2/(1.25)^4$ \eqref{eq_Mqq_AppendixD} for $qq\to qq$ \\
$|\mathcal{M}_{q\bar{q}}|^2/(1.25)^4$ \eqref{eq_Mqqbar_AppendixD} for $q\bar{q}\to q\bar{q}$}
& \makecell{$g(T)$ \vspace{0.2em}\\\eqref{eq:g_fit}}\\\hline 
\end{tblr}
\caption{Parameters used in Fig. \ref{eta/s_averaged}.}
 \label{table:paramters_fig5}
\end{table}


\twocolumngrid

\bibliography{Shear_viscosity_of_a_binary_mixture/biblio}

\end{document}